\documentclass[12pt,preprint]{aastex}

 \pdfoutput=1

\shorttitle{Morphology and Interaction of three CMEs}
\shortauthors{Mishra et al.}
\usepackage{pdflscape}
\usepackage{rotating}
\usepackage{epstopdf}
\newcommand{\aig}{    {\it Ann. Int. Geophys. }}

\begin{document}

\title{Morphological and Kinematic Evolution of Three Interacting Coronal Mass Ejections of 2011 February 13-15}


\author{Wageesh Mishra and Nandita Srivastava}
\affil{Udaipur Solar Observatory, Physical Research Laboratory, P. O. Box 198, Badi Road,
Udaipur 313001, India}

\email{wageesh@prl.res.in}

\begin{abstract}
During 2011 February 13 to 15, three Earth-directed CMEs launched in successively were recorded as limb CMEs by coronagraphs (COR) of STEREO. These CMEs provided an opportunity to study their geometrical and kinematic evolution from multiple vantage points. In this paper, we examine the differences in geometrical evolution of slow and fast speed CMEs during their propagation in the heliosphere. We also study their interaction and collision using \textit{STEREO}/SECCHI COR and Heliospheric Imager (HI) observations. We have  found evidence of interaction and collision between the CMEs of February 15 and 14 in COR2 and HI1 FOV, respectively, while the CME of February 14 caught the CME of February 13 in HI2 FOV. By estimating the true mass of these CMEs and using their pre and post-collision dynamics, the momentum and energy exchange between them during collision phase are studied. We classify the nature of observed collision between CME of February 14 and 15 as inelastic, reaching close to elastic regime. Relating imaging observations with the in situ measurements, we find that the CMEs move adjacent to each other after their collision in the heliosphere and are recognized as distinct structures in in situ observations by WIND spacecraft at L1. Our results highlight the significance of HI observations in studying CME-CME collision for the purpose of improved space weather forecasting.  

\end{abstract} 

\keywords{Sun: coronal mass ejections, Sun: heliosphere, shock waves}

\section{Introduction}
\label{Intro}
CMEs are frequent expulsions of massive magnetized plasma from the solar corona in to the heliopshere. If the CMEs are directed towards the Earth and have enhanced southward magnetic field, they can result in severe geomagnetic storms \citep{Dungey1961,Gosling1990,Echer2008}. The typical transit time of CMEs from the Sun to the Earth is between 1 to 4 days and the number of CMEs launched from the Sun is about 3 per day at maximum solar activity \citep{StCyr2000}. Therefore, interaction of CMEs in the heliosphere is expected to be more frequent near the solar maximum. The possibility of CME-CME interaction has been reported much earlier by analyzing in situ observations of CMEs by Pioneer 9 spacecraft \citep{Intriligator1976}. Also, \citet{Burlaga1987} presented an additional case of CME-CME interaction in the heliosphere using in situ observations of twin Helios spacecraft. They showed that compound streams are formed due to such interactions which have amplified parameters responsible for producing major geomagnetic storms. Using wide field-of-view (FOV) coronagraphic observations of \textit{Large Angle Spectrometric COronagraph (LASCO)} \citep{Brueckner1995} on-board \textit{SOlar and Heliospheric observatory (SOHO)} and long wavelength radio observations, \citet{Gopalswamy2001} first time provided the evidence for CME-CME interaction. \citet{Burlaga2002} identified a set of successive halo CMEs directed towards the Earth and found that they appeared as complex ejecta near 1 AU \citep{Burlaga2001}. They inferred that these CMEs, launched successively, merged en route from the Sun to the Earth and formed complex ejecta in which identity of individual CMEs was lost.

CME-CME interactions are also important as they can result in extended period of enhanced southward magnetic field which can cause intense geomagnetic storms \citep{Farrugia2004,Farrugia2006}. Also, they help to understand the collisions between large scale magnetized plasmoids and hence various plasma processes involved. Also, if a shock from a following CME penetrates a preceding CME, it provides a unique opportunity to study the evolution of the shock strength and structure and its effect on preceding CME plasma parameters \citep{Lugaz2005,Mostl2012,Liu2012}. Since, estimating the arrival time of CMEs at the Earth is crucial for predicting space weather effects near the Earth and CME-CME interactions are responsible for changing the dynamics of interacting CMEs in the heliosphere, therefore, such interactions need to be examined in detail. Further, reconnection between magnetic flux ropes of CMEs can be explored by studying cases of CME-CME interactions \citep{Gopalswamy2001,Wang2003} which are also known to lead to solar energetic particles (SEPs) events \citep{Gopalswamy2002}.

Prior to the launch of \textit{Solar TErrestrial RElations Observatory (STEREO)} \citep{Kaiser2008}, CME interaction study was limited to analyzing imaging observations near the Sun and in situ observations near the Earth and simulations studies \citep{Vandas1997,Odstrcil2003,Lugaz2005}. After the launch of \textit{STEREO} in 2006, its Sun Earth Connection Coronal and Heliospheric Investigation (SECCHI) \citep{Howard2008} package, which consists of remote sensing instruments, is capable of imaging a CME from its lift-off in the corona up to the Earth and beyond. It also enables us to witness CME-CME interaction in the heliosphere. Using Heliospheric Imagers (HIs) observations, studies of CME interaction has increased significantly, e.g. interacting CMEs of 2008 November 2 \citep{Shen2012}, of 2010 May 23-24 CMEs \citep{Lugaz2012} and the extensively studied CMEs of 2010 August 1 \citep{Harrison2012,Liu2012,Mostl2012,Temmer2012}.

In the present work, we have selected three Earth-directed interacting CMEs launched during 2011 February 13 - 15 from NOAA AR 11158 when the twin \textit{STEREO} spacecraft were separated by approximately 180$^\circ$. The interaction of these CMEs has been studied based on imaging and in situ observations from \textit{STEREO} and WIND spacecraft, respectively. Such a study is required to improve our understanding about nature and consequences of interaction. These interacting CMEs have also been studied earlier, e.g. by \citet{Maricic2014} and \citet{Temmer2014}.

 \citet{Maricic2014} used the plane of sky (POS) approximation and Harmonic Mean (HM) method \citep{Lugaz2009} to convert the derived elongation-time profiles from single \textit{STEREO} spacecraft to distance-time profiles and then estimated the arrival time of CMEs at L1. Their approach seems to be less reliable at larger elongation where the direction of propagation and structure of CME play a crucial role. Further, \citet{Maricic2014} could track a CME feature only up to small elongations ($\approx$ 25$\arcdeg$). However, in our analysis we have constructed \textit{J}-maps \citep{Sheeley1999,Davies2009} which allow us to follow the CMEs to significantly greater elongations ($\approx$ 45$\arcdeg$). Previous studies have shown that tracking of CMEs to larger elongation using \textit{J}-maps and subsequent  stereoscopic reconstruction give more precise kinematics and estimates of arrival time of CMEs than employing single spacecraft reconstruction methods \citep{Williams2009,Liu2010,Lugaz2010,Mishra2013,Colaninno2013,Mishra2014}. In our study, we have applied the stereoscopic methods (SSSE; \citealt{Davies2013}, TAS; \citealt{Lugaz2010.apj} and GT; \citealt{Liu2010}) which apart from dynamics also yield the time-variations of direction of propagation of CMEs. \citet{Lugaz2012} have found a change in the longitudinal direction of propagation of CMEs during their interaction. Such a deflection is of prime concern for predicting CME arrival time at Earth and to understand the collision dynamics of CMEs, which are the main objectives of the present paper. We have estimated the kinematics of overall CME structure using Graduated Cylindrical Shell (GCS) model \citep{Thernisien2009} in COR2 FOV while \citet{Maricic2014} estimated the kinematics of a single tracked feature. The kinematics of overall CME structure in COR2 FOV is helpful for determining the probability of collision of CMEs beyond COR2 FOV. \citet{Temmer2014} studied the interaction of February 14-15 CMEs corresponding to different position angles measured over entire latitudinal extent of these CMEs. In this context, the present study is important as it also focuses on understanding the nature of collision by estimating momentum and energy exchange during collision phase of the CMEs.

The unique positioning of \textit{STEREO} spacecraft, first time since its launch, enticed us to do an additional study about the geometrical evolution of these Earth-directed CMEs in COR2 FOV from identical multiple viewpoints. The location of active region (S20E04 to S20W12 during February 13-15) for these CMEs allowed its SECCHI/COR coronagraph to observe these Earth-directed CMEs at the limb (i.e. plane orthogonal to the Sun-STEREO line) contrary to \textit{SOHO}/LASCO observations which always record such CMEs as halos. In this scenario, the CME observations are least affected by the projection effects in both SECCHI/COR-A and B FOV and hence crucial parameters, i.e. widths, speeds etc. that define the geo-effectiveness of CMEs can be determined with a reasonable accuracy. Morphological studies have been carried out earlier, assuming either a cone or ice cream cone model for CMEs for estimating the true angular width, central position angle, radial speed and acceleration of halo CMEs \citep{Zhao2002,Michalek2003,Xie2004,Xue2005}. Also, \citet{Howard1982,Fisher1984} suggested that the geometrical properties of CMEs can be described by cone model which can be used to estimate their mass. All the cone models assume that angular width of CMEs remain constant beyond a few solar radii as they propagate through the solar corona. \citet{Vrsnak2010,Vrsnak2013} also assumed  a constant cone angular width of a CME for developing the drag based model (DBM) of propagation of CMEs. {Our study of the morphological evolution of the selected CMEs, is expected to provide results that can help to refine the cone model by  incorporating the possible variation in angular width of CMEs corresponding to their different speed, i.e., slower, comparable and faster than the ambient solar wind speed.

In Section~\ref{MorphoCOR}, we present the morphological evolution of CMEs. In Section~\ref{Evolution}, the kinematics and interaction of CMEs in the heliosphere are discussed. In Section~\ref{CompWidth}, the angular widths of CMEs determined from 2D images are compared with the angular widths derived from the GCS model. In Section~\ref{Collision}, we focus on the nature of collision and estimate the energy and momentum transfer during collision of CMEs. In Section~\ref{Insitu}, in situ observations of CMEs are described. Section~\ref{Arrtime} \& ~\ref{Geomag} describes the arrival of CMEs at L1 and their geomagnetic response, respectively.  The main results of the present study are discussed in Section~\ref{ResDis}.

\section{Remote Sensing Observations and Analysis}
\label{RemObs}

In this section, we summarize the imaging observations of CMEs during 2011 February 13 to 15 taken by \textit{STEREO}/SECCHI package which consists of five telescopes, namely one Extreme Ultraviolet Imager (EUVI), two coronagraphs (COR1 and COR2) and two Heliospheric Imagers (HI1 and HI2). Unlike to COR, the HI camera is off-centered from the Sun center and can observe CMEs from the outer edge of FOV of COR2 up to the Earth and beyond \citep{Eyles2009}. At the time of observations of CMEs presented in this study, \textit{STEREO-A} was $\approx$ 87$^{\circ}$ West and \textit{STEREO-B} was $\approx$ 94$^{\circ}$ East of the Earth. They were approximately in ecliptic plane with 0.96 AU and 1.0 AU distance from the Sun. 

CME of February 13 (hereinafter, CME1) was observed by \textit{SOHO}/LASCO-C2 at 18:36 UT on 2011 February 13 as a faint partial halo CME with angular width of 276$^{\circ}$. In SECCHI/COR1-A and B, this CME appeared at 17:45 UT in SE and SW quadrant, respectively. The CME of February 14 (hereinafter, CME2) was first recorded by \textit{SOHO}/LASCO-C2 at 18:24 UT on 2011 February 14 as a halo CME. The CME2 appeared in SECCHI/COR1-A and B at 17:45 UT at the East and West limb, respectively. In SOHO/LASCO-C2 FOV, CME of February 15 (hereinafter, CME3) was first observed at 02:24 UT on 2011 February 15 as a halo CME. In SECCHI/COR1-A and B images, the CME3 was first observed at 02:05 UT at the East and West limb respectively.

\subsection{Morphological Evolution of CMEs in COR Field-Of-View}
\label{MorphoCOR}
 
We measured the geometrical properties (e.g. cone angle) of the selected CMEs by analyzing the SECCHI/COR2 images. Our aim is to study the deviation of CMEs from the ideal cone model for CMEs with different speeds. We based our analysis on the concept that slow and fast speed CMEs interact with solar wind differently and hence, deviation of each from the cone model might be different. In the present study, we did not use COR1 images, as near the Sun, within a few solar radii, magnetic forces are dominant and also, CMEs are not fully developed. We excluded the CME1 for morphological study because it was very faint in COR2 FOV. We selected CME2 and CME3 for the morphological analysis which had different speeds in COR2 FOV.  

We began with the ice-cream model of \citet{Xue2005} which considers the shape of CMEs as a symmetrical cone combined with a sphere. As per this model, the apex of a cone and center of sphere are both located at the center of the Sun and CMEs move radially outward having constant cone angular width beyond few solar radii from the Sun. We measured the cone angular width of CMEs using COR2 images and estimated the cone area, i.e. $A$ = $\pi$$r^{2}$$\Theta$/360, where $\Theta$ is cone angular width in degrees and $r$ is the radius of the sphere which is equal to the distance between the front edge of CME and the center of the Sun. For the estimation of $r$ and $\Theta$, we avoided the use of coordinate conversion as we deal with CMEs approximately perpendicular to the camera of \textit{STEREO} spacecraft. Hence, the estimated area using above equation, is the area of the CME as it appears from sideways (perpendicular to its motion).

In order to calculate the cone area as described above, we processed the SECCHI/COR2 images of CME2 and CME3 and then subtracted a background image from them. Further, we enclosed the CME area by manual clicking and joining the points on CME boundary. We also used few initial points on each side of the CME flank close to the coronagraph occulter to get a cone model fit for each structure. These points are used to estimate the  position angle at both flanks (near the apex of cone) of CMEs. The difference in position angle at both flanks is the 2D angular width of CME. In the top panel of the Figure~\ref{CME2_contour}, evolution of slow speed CME2 as observed in COR2-A images is shown with overlaid contour enclosing the entire CME and over-plotted lines denote limiting position angle at both the flanks of CME. We repeated this analysis for fast speed CME3, and its appearance in COR2-A FOV is shown in bottom panel of Figure~\ref{CME2_contour}.

The estimated 2D cone angular width for CME2 and CME3 in COR2 FOV is shown in Figure~\ref{theta_CMEs}. From the figure, it is evident that slow speed CME2 has nearly constant (between 60$^{\circ}$ to 57$^{\circ}$) 2D angular width in the COR2 FOV. For the fast speed CME3, its 2D angular width was $\approx$ 80$^{\circ}$ in the begining which then decreased to 62$^{\circ}$ as it crossed the outer edge of COR2 FOV. From the contour in Figure~\ref{CME2_contour} (top panel), it can be seen that CME2 followed the cone model and a slight spill of CME2 on the upper edge is compensated by a gap with the lower edge. For the fast speed CME3 (bottom panel of Figure~\ref{CME2_contour}), we noticed a significant spill on both sides (upper and lower  edge) which  increased with time in COR2-A FOV. The appearance and the variations of angular width of CME2 and CME3 in COR2-B images are same as in COR2-A.

To calculate the ice cream cone model area (cone area) for both CMEs, we marked a point along leading edge on each image, the distance of which from the center of the Sun gave the radius of the sphere located on the cone like CME. Since, some part of the CME is blocked by the occulter in all the images, therefore the area of CME blocked by the occulter has been subtracted from the sector (cone) area in order to compare it with the actual contour area of the CME. We also calculated the actual area enclosed by the CME contour (contour area). For convenience, both cone and contour areas have been measured in units of pixel$^{2}$. This is not of much concern as it is the difference in the actual and the sector area that we are interested in. In the top panels of Figure~\ref{CME2_area} and ~\ref{CME3_area}, blue curve represents the cone area (i.e. area obtained by approximating the CME with a cone model) and red curve represents the actual contour area. From these figures, we note that for both CME2 and CME3, time-variation of cone and contour area show a parabolic pattern in COR2-A and B FOV, which implies that $A$ is proportional to $r^{2}$. Therefore, we consider that both fast and slow CMEs follow the cone model to a certain extent.

From the top panel of Figure~\ref{CME2_area}, we find that cone area is larger than the contour area and both increased with time. In the bottom panels, we see that the difference in cone area and contour area is positive and increased as the CME2 propagated in the outer corona. For CME3, we can see (in top panel of Figure~\ref{CME3_area}) that the line representing the cone and contour area intersect one another in both COR2-A and B FOV. In the bottom panels of this figure, the difference in cone and contour area decreases from positive (2.8 $\times$ 10$^{4}$ pixel$^{2}$ in COR2-A and 2.4 $\times$ 10$^{4}$ pixel$^{2}$ in COR2-B) to negative values and remains negative (-2.4 $\times$ 10$^{4}$ pixel$^{2}$ in COR2-A and -3.9 $\times$ 10$^{4}$ pixel$^{2}$ in COR2-B) as the CME3 propagated through the outer corona. This is suggestive that at a certain height during its propagation in COR2 FOV, the contour area becomes larger than the estimated cone area. These findings indicate dissimilar morphological evolution for slow and fast speed CMEs in the corona.

\subsection{Kinematic Evolution and Interaction of CMEs in the Heliosphere} 
\label{Evolution}
\subsubsection{3D Reconstruction in COR2 Field-Of-View}
\label{ReconsCOR}

The launch of CME1, CME2 and CME3 from the same active region in quick succession indicates a possibility of their interaction as they move out from the Sun in to the heliosphere. To estimate the true kinematics of these CMEs, we have carried out the 3D reconstruction of CMEs using Graduated Cylindrical Shell (GCS) model developed by \citet{Thernisien2009}. We apply the GCS model to contemporaneous images from SECCHI/COR2-B, SOHO/LASCO and SECCHI/COR2-A. Before applying the model, the total brightness images were processed and then a pre event image was subtracted from a sequence of images to which the GCS model was applied. The images of CME1, CME2 and CME3 overlaid with the fitted GCS wireframed contour (hollow croissant) are shown in Figure~\ref{CME123_FM}.

The true kinematics estimated for these CMEs in COR2 FOV are shown in Figure~\ref{CME123_3DFM}. As the CME1 was faint and non-structured, GCS model fitting could be done only for 3 consecutive images in COR2 FOV. The estimated longitudes ($\phi$) for CME1, CME2 and CME3 at their last estimated height of 8.2 \textit{R}$_{\odot}$, 10.1 \textit{R}$_{\odot}$ and 11.1 \textit{R}$_{\odot}$ are -2$^{\circ}$, 6$^{\circ}$ and -3$^{\circ}$, respectively. The estimated latitudes ($\theta$) at these heights are -6$^{\circ}$, 4$^{\circ}$, -11$^{\circ}$ for CME1, CME2 and CME3, respectively. The estimated 3D speed at their last estimated heights for CME1 (February 13, 20:54 UT), CME2 (February 14, 22:24 UT) and CME3 (February 15, 03:54 UT) is found to be 618 km s$^{-1}$, 418 km s$^{-1}$ and 581 km s$^{-1}$, respectively. From the kinematics plot (Figure~\ref{CME123_3DFM}), it is clear that CME3 was faster than the preceding CME2 and  headed approximately in the same direction towards the Earth. Moreover, the launch of CME3 preceded that of CME2 by $\approx$ 9 hr, therefore, it is expected that these CMEs would interact at a certain distance in the heliosphere. Since the direction of propagation of CME1 and CME2 was also same, there exists a possibility of interaction between CME1 and CME2 in case if CME1 decelerates and CME2 accelerates beyond the estimated height in COR2 FOV. From the 3D reconstruction in COR2 FOV, we found that the speed of CME3 decreased very quickly from 1100 km s$^{-1}$ at 6 \textit{R}$_{\odot}$ to 580 km s$^{-1}$ at 11\textit{R}$_{\odot}$ during 02:39 UT to 03:54 UT on 2011 February 15. A quick deceleration of fast CME3 within 1.5 hr is most likely due to the interaction between CME2 and CME3. The terminology `interaction' and `collision' as used in this paper stand for two different sense. By `interaction', we mean that one CME causes deceleration or acceleration of another, although no obvious signature of merging of propagation tracks of features corresponding to the two CMEs is noticed in \textit{J}-maps.  The `collision', here is referred as the phase during which the tracked features of two CMEs moving with different speeds come in close contact with each other until they achieve an approximately equal speed or their trend of acceleration is reversed or they get separated from each other. The fast deceleration of CME3 from the beginning of the COR2 FOV may occur due to various possibilities. It may be either due to the presence of dense material of the preceding CME2 or due to the decrease of magnetic driving forces of CME3 or due to the overlying curved magnetic field lines of the preceding CME2 which can act as a magnetic barrier for CME3 \citep{Temmer2008,Temmer2010,Temmer2012}.

\subsubsection{Comparison of angular widths of CMEs derived from GCS model}
\label{CompWidth}

As discussed in Section~\ref{MorphoCOR}, we have estimated the cone angular width of CMEs using 2D COR2 images, it appears relevant to compare this to angular width determined from the GCS model of 3D reconstruction. Using the GCS 3D reconstruction technique, apart from the kinematics of CMEs (explained in Section~\ref{ReconsCOR}), we also obtained the aspect ratio ($\kappa$) of GCS model for CME1, CME2 and CME3 as 0.25, 0.28 and 0.37, respectively, at the last point of estimated distance in COR2 FOV. Also, we found the tilt angle ($\gamma$) around the axis of symmetry of the model as 7$\arcdeg$, -8$\arcdeg$ and 25$\arcdeg$ for CME1, CME2 and CME3, respectively. The positive (negative) value of the tilt shows that the rotation is anticlockwise (clockwise) out of the ecliptic plane. The angular width (2$\alpha$) between the legs of the GCS model is 34$\arcdeg$, 64$\arcdeg$ and 36$\arcdeg$ for CME1, CME2 and CME3, respectively. These values are in agreement (within $\pm$ 10\%) with the values obtained by \citet{Temmer2014}. It is to be noted that measured 2D angular width of CME depends on the orientation of the GCS flux ropes. For ecliptic orientation of the flux ropes, i.e., $\gamma$ = 0$\arcdeg$, the angular width of CME seen in 2D images is equal to the 3D edge-on angular width ($\omega$$_{EO}$ = 2$\delta$) of GCS model, where $\delta$ = $\arcsin$($\kappa$). For $\gamma$ = 90$\arcdeg$, the measured 2D width is equal to 3D face-on angular width ($\omega$$_{FO}$ = 2$\alpha$+2$\delta$) of the GCS modeled CME.

We converted the GCS modeled 3D width to 2D angular width for CME2 using the expression, $\omega$$_{2D}$ = $\omega$$_{EO}$$\cos$($\gamma$)+$\omega$$_{FO}$$\sin$($\gamma$), and find that CME2 has approximately constant 2D angular width in COR2 FOV. We find that the fast speed CME3 has $\gamma$ = 21$\arcdeg$, $\kappa$=0.40 and $\alpha$ = 16$\arcdeg$ in the beginning of the COR2 FOV while it has $\gamma$ = 21$\arcdeg$, $\kappa$=0.31 and $\alpha$ = 18$\arcdeg$ at the last measured point in COR2 FOV. Hence, we find that as  CME3 propagates further in COR2 FOV, its 2D angular width (derived using GCS modeled 3D width) decreases from 77$\arcdeg$ to 63$\arcdeg$. These findings are in accordance with the observed 2D angular width of CME2 and CME3 (Figure~\ref{theta_CMEs}). The possibility of rotation of CMEs have been discussed theoretically (\citealt{Lynch2009}, and references therein) and reported in low corona observations \citep{Lynch2010}. Such changes in the 2D measured angular width is also possible due to rotation (change in CME orientation, i.e. tilt, angle) of fast speed CME towards or away from the equator as shown by \citet{Yurchyshyn2009} who suggest higher rotation rate for a faster CME \citep{Lynch2010,Poomvises2010}. However, we emphasize that based on the GCS modeling, we could not infer any noticeable rotation (change in $\gamma$) or deflection (change in $\phi$ in Figure~\ref{CME123_3DFM}) in the COR2 FOV for the selected CMEs. The uncertainties involved in the estimation of 3D and observed 2D angular widths are discussed in Section~\ref{ResDis}.

\citet{Vourlidas2011} reported that despite the rapid rotation of CMEs there is no significant projection effects (change in angular width) in any single coronagraphic observations. They showed that the projected (2D) angular width of CME altered by only 10$\arcdeg$ between 2 to 15 \textit{R}$_{\odot}$ while CME rotated by 60$\arcdeg$ over the same height range. We acknowledge the error in the measurements of tilt angle in our study, but it must be noted that a rotation of $\approx$ 40$\arcdeg$ within 6 hr is required for the observed large variations in the 2D angular width of CME3, which is indeed not found in our analysis. Therefore, we consider that observed decrease in the angular width of CME3 is not because of its rotation, but may be due to its interaction with solar wind or dense material from preceding CME2.

\subsubsection{Reconstruction of CMEs in HI FOV} 
\label{Reconstruction HI}

Based on the kinematics observed close to the Sun, i.e. using COR observations, we consider the possibility that these Earth-directed CMEs have a chance of interaction and therefore we estimated their kinematics in HI FOV. \setlength{\emergencystretch}{2em}{We used the long term background subtracted Level 2 data for HI observations taken from UKSSDC \url{(http://www.ukssdc.ac.uk/solar/stereo/data.html)}. We examined the base difference images in HI FOV to notice any density depletion or enhancement due to CME. Prior to this step, HI image pair was aligned to prevent the stellar contribution in difference images. Further, we notice that CME3 approached and meet with CME2 in the HI1 FOV. In this collision, leading edge of CME3 flattened significantly. This observation  motivated us, to investigate the pre and post-collision kinematics of CMEs. Since in HI FOV, CMEs become faint and tracking of the features out to larger distances invoke the uncertainties. Therefore, we track CMEs in the heliosphere by constructing the time-elongation map (\textit{J}-map) \citep{Davies2009}, originally developed by \citet{Sheeley1999} and as described in detail in Section 3.1.1 of \citet{Mishra2013}.

The constructed \textit{J}-map in the ecliptic plane for these CMEs in HI-A and B FOV are shown in Figure~\ref{J-maps}. By tracking the bright leading fronts manually, we derive the elongation-time profiles for all the three CMEs. In this figure, derived elongation for the outward moving CMEs are overplotted with dotted colored lines. The CME1 is very faint and could be tracked out to $\approx$ 13$^{\circ}$ in the \textit{STEREO-A} and \textit{B} \textit{J}-maps. However, CME2 and CME3 could be tracked out to  44$^{\circ}$ and 46$^{\circ}$ in \textit{STEREO-A} \textit{J}-maps, respectively, and out to $\approx$ 42$^{\circ}$ in \textit{STEREO-B} \textit{J}-maps. \textit{J}-maps also show that the bright tracks of CME2 and CME3 approach close to each other suggesting their possible collision in HI FOV.

Various stereoscopic reconstruction methods have been developed to estimate the kinematics of CMEs using SECCHI/HI images \citep{Liu2010,Lugaz2010.apj,Davies2013}. The selected CMEs in our study have a cone angular width of $\approx$ 60$^\circ$, therefore, it is preferable to use those reconstruction methods which take into account the geometry of CMEs with similar angular width. Keeping these points in mind, we implemented the stereoscopic self-similar expansion (SSSE) \citep{Davies2013} method on the derived time-elongation profiles for all three CMEs to estimate their kinematics. While applying this method, we fixed the CME's cross-sectional angular half width subtended at the Sun equal to 30$^{\circ}$. Using the SSSE method, for all the three CMEs the kinematics, i.e. estimated height, direction and speed, were obtained (Figure~\ref{CME123_SSSE}). The speed was derived from the adjacent distance points using numerical differentiation with three point Lagrange interpolation  and therefore have systematic fluctuations. Estimating the speed in this way can provide short time variations in CME speed during its interaction with solar wind or other plasma density structures in the solar wind. On the other hand, the smoothed speed can also be derived if the estimated distance is fitted overall into an appropriate polynomial, but the information about variations of speed will be lost. Also, by fitting a polynomial for the derived fluctuating speeds, the speed can be shown with minimal fluctuations. Therefore, we have made a compromise and fitted the estimated distance during each 5 hour interval into a first order polynomial and derived the speed which is shown with horizontal solid lines in the bottom panel of Figure~ \ref{CME123_SSSE}. The error bars for the estimated parameters are also shown in this figure with vertical solid lines at each data point. The detailed procedure of the estimation of error bars is described in Section~\ref{CompKinem}. Propagation direction of the CME2 and CME3 seem to follow the same trajectory and are approximately Earth-directed. However, the unexpected variations in the direction of propagation of both CMEs were noticed which are discussed in Section~\ref{ResDis}. In Figure~\ref{CME123_SSSE}, we noticed a jump in the speed of CME2 and CME3 at 08:25 UT on 2011 February 15. Within 18 hr, after an increase in speed is observed, the speed of the CME2 increases from about 300 km s$^{-1}$ to 600 km s$^{-1}$. During this time the speed of CME3 decreased from about 525 km s$^{-1}$ to 400 km s$^{-1}$. Later, both CMEs achieved a similar speed of $\approx$ 500 km s$^{-1}$. Such a finding of acceleration of one and deceleration of another CME, supports a possible collision between CME2 and CME3. The collision phase is shown in the top and bottom panels of Figure~\ref{CME123_SSSE} (region between the two dashed vertical lines, from the left). After the collision, we find that both the CMEs move in close contact with each other.

The strong deceleration of CME3 observed prior to the merging of the bright tracks (enhanced density front of CMEs) in J-maps, suggests possible interaction of CME3 with CME2. This is possible as we track the leading front of the CMEs using \textit{J}-maps, the trailing edge of CME2 can cause an obstacle for CME3 leading front, much earlier depending on its spatial scale. From the observed timings, it is clear that interaction of CME3 with CME2 had started $\approx$ 5 hr prior to their collision in HI FOV. Our analysis also shows that the leading front of CME3 reflects the effect of interaction (i.e. strong deceleration) at 6 \textit{R}$_{\odot}$ while leading edge of CME2 shows this effect (i.e. acceleration) at 28 \textit{R}$_{\odot}$. Therefore, the force acting on the trailing edge of CME2 takes approximately $\approx$ 5.7 hr to reach the leading front of this CME. Based on these values, the propagation speed of disturbance responsible for acceleration of leading front of CME2 should be $\approx$ 750 km s$^{-1}$. From the Radio and Plasma Wave Experiment (WAVES) \citep{Bougeret1995} on board WIND spacecraft, we noticed a type II burst during 02:10 - 07:00 UT in 16000 - 400 KHz range. Such radio bursts provide information on CME driven shock \citep{Gopalswamy2000a}. This shock is associated with the fast speed CME3. The average shock cone angle ($\approx$ 100$^\circ$) as seen from the Sun is significantly greater than average angular size ($\approx$ 45$^\circ$) of any CME \citep{Schwenn2006}. It is likely that this shock traveled across CME2. Therefore, the acceleration of CME2, observed in HI1 FOV, may be due to the combined effect of the shock and leading front of CME3. As previously mentioned, CME1 was very faint, and its kinematics could be estimated up to 46 \textit{R}$_{\odot}$ only. Based on the linear extrapolation of the height-time curve of CME1 and CME2, we infer that they should meet each other at 144 \textit{R}$_{\odot}$ at 01:40 UT on 2011 February 17.

\subsubsection{Comparison of Kinematics Derived from Other Stereoscopic Methods} 
\label{CompKinem}
To examine the range of uncertainties in the estimated kinematics of the CMEs of 2011 February 13-15, by implementing SSSE method, we applied another stereoscopic method, viz. Tangent to a sphere (TAS) \citep{Lugaz2010.apj} method to all three CMEs. Based on the estimated kinematics, we infer similar acceleration of CME2 and deceleration of CME3, as obtained by implementing SSSE method.  Using TAS method, we found that the leading edge of CME3 caught the leading edge of CME2 at 26 \textit{R}$_{\odot}$ at 08:24 UT on 15 February. Based on linear extrapolation of the height-time profile of CME2 and CME1, we inferred that CME2 would have reached CME1 at 157 \textit{R}$_{\odot}$ at 03:35 UT on 2011 February 17. We also implemented the Geometric Triangulation (GT) \citep{Liu2010} method using the derived elongation-time profiles of tracked features of these CMEs to estimate their kinematics. On using GT method, we find similar results as obtained from TAS and SSSE methods. Based on the estimated kinematics from GT method, we note that the leading edge of CME3 caught the leading edge of CME2 at 24 \textit{R}$_{\odot}$ at 07:10 UT on February 15. The linear extrapolation of height-time for CME1 and CME2 suggest their interaction at 138 \textit{R}$_{\odot}$ on 20:24 UT on February 17.

The kinematics derived from SSSE method is shown in Figure~\ref{CME123_SSSE} and cognizance of the involved uncertainties  is important. However, the actual uncertainties in the derived kinematics owe to several factors (geometry, elongation measurements, Thomson scattering, line of sight integration effect, breakdown of assumptions in the method itself) and its quantification is extremely difficult. \citet{Davies2013} have shown that GT and TAS methods are special cases of SSSE method corresponding to two extreme cross-sectional extent (geometry) of CME, i.e., corresponding to angular half width of $\lambda$ = 0$\arcdeg$ and $\lambda$ = 90$\arcdeg$, respectively. We estimated the uncertainties due to consideration of different geometry in each of the three implemented stereoscopic techniques (GT, TAS and SSSE). Such uncertainties are shown with error bars with vertical solid lines in Figure~\ref{CME123_SSSE}. We estimated the absolute difference between kinematics values derived from SSSE and GT method and display it as a vertical lower error (lower segment of error bars). Similarly, the absolute difference between kinematics values from SSSE method and TAS method is displayed as vertical upper error. From Figure~\ref{CME123_SSSE}, we notice that the results from all three methods are in reasonable agreement.

Further, we attempted to examine the contribution of errors in the kinematics due to limited accuracy in tracking (i.e. elongation measurements) of a selected feature. Following the error analysis approach of \citet{Liu2010.722}, we consider an uncertainty of 10 pixels in elongation measurements from both STEREO viewpoints which correspond to elongation uncertainty of 0.04$\arcdeg$, 0.2$\arcdeg$ and 0.7$\arcdeg$ in COR2, HI1 and HI2 FOV, respectively. This leads to an uncertainly of 0.20 - 0.35 \textit{R}$_{\odot}$, 0.21 - 0.75 \textit{R}$_{\odot}$, and 0.19 - 0.74 \textit{R}$_{\odot}$ in the estimated distance for CME1, CME2 and CME3, respectively. Such small uncertainties in the distance is expected to result in error of less than $\approx$ 100 km s$^{-1}$ in speed. However, similar elongation uncertainty lead to crucially larger uncertainty in the estimated direction of propagation of CMEs when they are close to entrance of HI1 FOV, where singularity occurs. The occurrence of singularity is described in earlier studies \citep{Liu2011,Mishra2013,Mishra2014}. The estimated propagation direction of CMEs from GT method are shown in Figure~\ref{CME123_GTErr} in which vertical lines at each data point show the uncertainty in the direction.

Based on the aforementioned error estimation for the selected CMEs in our study, we find that the uncertainties in the estimated kinematics from stereoscopic methods owe mostly due to errors in elongation measurements rather than geometry. Due to the large separation between the two STEREO viewpoints and consequently occurrence of singularity, small observational errors in the elongation measurements yield significantly larger errors in the kinematics (especially in the direction), irrespective of the geometry considered for the CMEs \citep{Davies2013,Mishra2014}.

\subsection{Energy, Momentum Exchange and Nature of Collision Between CME2 and CME3}
\label{Collision}

The dynamics and structure of CMEs are likely to change when they collide with one another, therefore, estimation  of 
post-collision kinematics is essential to achieve the goal of space weather prediction. As the CMEs are large scale magnetized plasmoids which interact with each other, it is worth investigating the nature of collision for CMEs which is expected to be different than the collision of gaseous bubbles with no internal magnetic field. In collision dynamics, the total momentum of colliding bodies is conserved irrespective of the nature of collision provided that external forces are absent.

We attempt to investigate the nature of collision for CME2 and CME3. As the CME3 follows the trajectory of CME2 before and also after the collision, we simply use the velocity derived from SSSE method to deal with the collision dynamics. Therefore, we did not take into account the 3D velocity components and intricate mathematics for determining the motion of centroid of colliding CMEs, as used in \citet{Shen2012}. We studied one-dimensional collision dynamics which is similar to the case of head-on collision for the interacting CMEs. We note that the start of collision phase (marked by the dashed vertical line) occurs at the instant when the speed of CME2 started to increase while the speed of CME3 started to decrease (Figure~\ref{CME123_SSSE}, bottom panel). The same trend of speeds for both the CMEs are maintained up to 18 hr where the collision phase ends. After the end stage of the collision phase, CME2 and CME3 show trend of deceleration and acceleration, respectively, towards a constant speed of 500 km s$^{-1}$. From the obtained velocity profiles (Figure~\ref{CME123_SSSE}), we notice that velocity of CME2 and CME3 before the collision are u$_{1}$ = 300 and u$_{2}$ = 525 km s$^{-1}$, respectively. After the collision and exchange of velocity, the velocity of CME2 and CME3 is found to be v$_{1}$ = 600 and v$_{2}$ = 400 km s$^{-1}$, respectively.  If the true mass of CME2 and CME3 be m$_{1}$ and m$_{2}$, respectively, then the conservation of momentum requires m$_{1}$u$_{1}$ + m$_{2}$u$_{2}$ = m$_{1}$v$_{1}$ + m$_{2}$v$_{2}$. To examine the momentum conservation for the case of colliding  CMEs, we need to calculate the true mass of both CMEs, which is discussed in the following section. 

\subsubsection{Estimation of True Mass of CMEs}
\label{TrueMass}

Historically, mass of a CME has been calculated using the POS approximation which resulted in a underestimated value \citep{Munro1979,Poland1981,Vourlidas2000}. We implemented the approach of \citet{Colaninno2009} and derived the true propagation direction and then the true mass of CMEs in COR2 FOV. Before applying this approach, base difference images were obtained following the procedure described in \citet{Vourlidas2000,Vourlidas2010,Bein2013}. To estimate the projected mass of CMEs in base difference COR2-A and B images, we selected a region of interest (ROI) which enclosed the full extent of a CME. The intensity at each pixel was then converted to the number of electrons at each pixel and then the mass per pixel was obtained. The total mass of CME was calculated by summing the mass at each pixel inside this ROI. In this way, we estimated the projected mass of CME, M$_{A}$ and M$_{B}$ from two viewpoints of \textit{STEREO-A} and \textit{B} in COR2 FOV.

According to \citet{Colaninno2009}, CME mass  M$_{A}$ and M$_{B}$ are expected to be equal as the same CME volume is observed from two different angles. Any difference between these two masses, it must be due to incorrect use of the  propagation angle in the Thomson scattering calculation. Based on this assumption, they derived an equation for true mass (M$_{T}$) as a function of projected mass and true direction of propagation of CME (see their equations 7 and 8). We used a slightly different approach to solve these equations, viz.
\begin{equation}
M_{A}/M_{B} = B_{e}(\theta_{A})/B_{e}(\theta_{A} + \Delta)
\end{equation}
where $\theta$$_{A}$ is the angle of direction of propagation of CME measured from the POS of \textit{STEREO-A}, B$_{e}$($\theta$$_{A}$) is the brightness of a single electron at an angular distance of $\theta$$_{A}$ from the POS and $\Delta$ is the summation of longitude of both \textit{STEREO-A} and \textit{STEREO-B} from the Sun-Earth line. Once we obtained the measured values of M$_{A}$ and M$_{B}$, we derived its ratio and calculate $\theta$$_{A}$. In this way, we obtained multiple values of $\theta$$_{A}$ which result in same value of the ratio of M$_{A}$ and M$_{B}$. The correct value of $\theta$$_{A}$ was found by visual inspection of CME images in COR FOV. Once we obtained the $\theta$$_{A}$, the true mass of CME was estimated using the equation (4) of \citet{Colaninno2009}. Here we must emphasize that estimation of true propagation direction of CMEs ($\theta$$_{A}$ ) using aforementioned approach has large errors if the value of $\Delta$ approaches 180$^\circ$. This is a severe limitation of the method of true mass estimation and arises because in such a scenario a CME from the Sun, despite its propagation in any direction (not only towards the Earth), will be measured at equal propagation angle from the POS of both spacecraft. Therefore, in principle both the estimated M$_{A}$ and M$_{B}$ should be exactly equal and any deviation (which is likely) will result in highly erroneous value of $\theta$$_{A}$, and consequently in the true mass of CME. Such a limitation has also been reported by \citet{Colaninno2009} for very small spacecraft separation angle. This implies that, an accurate propagation direction cannot be derived with this method unless we adjust the separation angle between STEREO spacecraft slightly. Hence, we use a slightly different value of $\Delta$ $\approx$ 160$^\circ$ for our case. By repeating our analysis several times for these CMEs, we noted that such a small change in $\Delta$ has negligibly small effect on CME mass. In our study, we have also determined the true mass using the 3D propagation direction obtained from another method (GCS forward fitting model) and found that these results are within $\approx$ 15\% of estimates from the method of  \citet{Colaninno2009}.

In the present work, we have estimated the true masses of CME2 and CME3 to understand their collision phase. For CME2 at a heliocentric distance of $\approx$ 10 \textit{R}$_{\odot}$, M$_{A}$ and M$_{B}$ was estimated as 5.30 $\times$ 10$^{12}$ kg and 4.38 $\times$ 10$^{12}$ kg, respectively and its propagation direction as 24$^{\circ}$ East from the Sun-Earth line. For the CME3 at $\approx$ 12 \textit{R}$_{\odot}$, M$_{A}$ and M$_{B}$ was estimated as 4.56 $\times$ 10$^{12}$ kg and 4.77 $\times$ 10$^{12}$ kg, respectively and its propagation direction as 30$^{\circ}$ East from the Sun-Earth line. The true mass of CME2 and CME3 is estimated as  $m_{1}$ = 5.40 $\times$ 10$^{12}$ kg and $m_{2}$ = 4.78 $\times$ 10$^{12}$ kg, respectively. We also noticed that mass of CMEs increased with distance from the Sun and we interpret such an increase in mass as an observational artifact due to emergence of CME material from behind the occulter of the coronagraphs, however the possibility of a small real increase in CME mass can not be ignored completely.

\subsubsection{Estimation of Coefficient of Restitution}
\label{CoefResti}

As per our calculations, the masses of CME were found to become constant after $\approx$ 10 \textit{R}$_{\odot}$ therefore, we assume that these masses remain constant before and after their collision in HI FOV. Combining the equation of conservation of momentum with coefficient of restitution, the velocities of CME2 and CME3 after the collision can be estimated theoretically ($v_{1th},v_{2th}$).

\begin{equation}  \label{v12th}
  v_{1th} = \frac{m_{1}u_{1} + m_{2}u_{2} + m_{2} e (u_{2} - u_{1})}{(m_{1} + {m_{2}})};
	v_{2th} = \frac{m_{1}u_{1} + m_{2}u_{2} + m_{1} e (u_{1} - u_{2})}{(m_{1} + {m_{2}})}
\end{equation}
 
where $e$ is the coefficient of restitution, $e$ = $v_{2}$ - $v_{1}$/$u_{1}$ - $u_{2}$ and signifies the nature of collision.

Using the velocity ($u_{1}$,$u_{2}$) = (300,525) km s$^{-1}$ and true mass values ($m_{1}$,$m_{2}$) = (5.40 $\times$ 10$^{12}$, 4.78 $\times$ 10$^{12}$) kg, we calculate a set of theoretical values of final velocity ($v_{1th}$,$v_{2th}$) after the collision of CMEs from equation~(\ref{v12th}) corresponding to a set of different values of coefficient of restitution ($e$). We define a parameter called variance, $\sigma = \sqrt{(v_{1th} - v_{1})^{2} + (v_{2th} - v_{2})^{2}}$. Considering the theoretically estimated final velocity from the equation~(\ref{v12th}) and variance ($\sigma$) values, one can obtain the most suitable value of $e$ corresponding to which the theoretically estimated final velocity ($v_{1th}$,$v_{2th}$) is found to be closest to the observed final velocity ($v_{1}$,$v_{2}$) of CMEs. This implies that the computed variance is minimum at this $e$ value.

We have estimated that total kinetic energy of the system before the collision as  9.01 $\times$ 10$^{23}$ joules. The individual kinetic energy of CME2 and CME3 is 2.43 $\times$ 10$^{23}$ joules and 6.58 $\times$ 10$^{23}$ joules, respectively. We note that, momentum of CME2 and CME3 is  1.6 $\times$ 10$^{18}$ N s and 2.5 $\times$ 10$^{18}$ N s, respectively, just before their observed collision. Hence, the total momentum of the system is equal to 4.13 $\times$ 10$^{18}$ N s. We consider ($v_{1}$,$v_{2}$) the estimated final velocity (from SSSE method) as (600,400) km s$^{-1}$  (Figure~\ref{CME123_SSSE}). We found that ($v_{1th}$,$v_{2th}$) = (495,304) km s$^{-1}$  and the minimum value of $\sigma$ is 142 corresponding to $e$ = 0.85. For this value of $e$, the momentum is found to be conserved and nature of collision is found to be in the inelastic regime. Such a collision resulted in a decrease of total kinetic energy of the system by 2\% of its value before the collision. If the coefficient of restitution is calculated using directly the measured values of velocity then $e$ is estimated as 0.89 which is approximately equal to what is calculated from using aforementioned theoretical approach.

To account for uncertainties in the results, we repeated our computation by taking an uncertainty of $\pm$ 100 km s$^{-1}$ in the estimated final velocity after the collision of CMEs. For example, if we use ($v_{1}$,$v_{2}$) = (700,500), then minimum value of $\sigma$ = 288 is found corresponding to $e$ = 0.80. The estimate for $\sigma$ is found to be minimum and is equal to 2.0 corresponding to $e$ = 0.90, when ($v_{1}$,$v_{2}$) = (500,300) km s$^{-1}$ is used and in this case ($v_{1th}$,$v_{2th}$) = (501,298) is obtained. It means that keeping the conservation of momentum as a necessary condition, the combination of 
($u_{1}$,$u_{2}$) = (300,525) km s$^{-1}$ and ($v_{1}$,$v_{2}$) = (500,300) km s$^{-1}$ with $e$ = 0.90 best suits the observed case of collision of CME2 and CME3. In this case, the total kinetic energy after the collision decreased by only 1.3 \%, the kinetic energy of CME2 increased by 177\%, and kinetic energy of CME3 decreased by 67\% of its value before the collision. It implies that observed collision is in the inelastic regime but closer to elastic regime. For this case, the momentum of CME2 increased by 68\% and momentum of CME3 decreased by 35\% of its value before their collision. Our analysis therefore shows that there is a huge transfer of momentum and kinetic energy during the collision phase of CMEs.

It is worth to check the effect of uncertainty in mass in the estimation of value of $e$ and hence on the estimation of nature of collision. We have estimated the true mass which is also uncertain  and difficult to quantify \citep{Colaninno2009}. However, a straightforward uncertainty arises from the assumption that CME structure lies in the plane of 3D propagation diretion of CME. \citet{Vourlidas2000} have shown that such a simplified assumption can cause the underestimation of CME mass by up to 15\%. Applying this error to estimated true mass of CME2 and CME3, their mass ratio ($m_{1}$/$m_{2}$ = 1.12) can range between 0.97 to 1.28. To examine the effect of larger uncertainties in the mass, we arbitrarily change the mass ratio between 0.5 to 3.0 in the step of 0.25 and repeat the aforementioned analysis (using equation~(\ref{v12th}) and calculating $\sigma$ value) to estimate the value of $e$ corresponding to each mass ratio. The variation of $e$ with mass ratio is shown in Figure~\ref{CME12_mass} corresponding to the observed final velocity ($v_{1}$,$v_{2}$) = (600,400) km s$^{-1}$ after collision of CMEs in our case. We have shown earlier that best suited final velocity of CMEs for our observed case of collision is ($v_{1}$,$v_{2}$) = (500,300) km s$^{-1}$, therefore, corresponding to this velocity also, the variation of $e$ with mass ratio is shown (Figure~\ref{CME12_mass}). In this figure, we have also plotted the estimated minimum variance corresponding to each obtained value of $e$. From this figure, it is evident that even if a large arbitrary mass ratio is considered, the nature of collision remains in the inelastic regime. It never reaches a completely inelastic ($e$ = 0), elastic ($e$ = 1) or super-elastic ($e$ $>$ 1) regime.

\section{In Situ Observations, Arrival Time and Geomagnetic Response of the Interacting CMEs}

\subsection{In Situ Observations}
\label{Insitu} 

We analyzed the WIND spacecraft plasma and magnetic field observations taken from CDAWeb $(http://cdaweb.gsfc.nasa.gov/)$. We attempted to identify the CMEs  based on criterion of \citet{Zurbuchen2006}. The variations in plasma and magnetic field parameters during 2011 February 17 at 20:00 UT to February 20 at 04:00 UT are shown in Figure~\ref{CME123_insitu}. The findings from in situ data analysis in our study are very similar to as reported by \citet{Maricic2014}. Since, we associate the remote observations to in situ observations and compare the arrival time of interacted CMEs, for sake of completeness, we briefly discuss in situ observations. In Figure~\ref{CME123_insitu}, the region marked as R1, R2 and R3 is associated with CME1, CME2 and CME3, respectively. In the region R3, the latitude and longitude of the magnetic field vector (from top, 6th and 7th panel of Figure~\ref{CME123_insitu}) seemed to rotate and plasma beta ($\beta$) was found to be less than one. Therefore, this region (R3) may be termed as a magnetic cloud (MC).

In region bounded between 09:52 UT and 10:37 UT on February 18 with third and fourth dashed lines, from the left, show sharp decrease in magnetic field strength, enhanced temperature and flow speed, as well as sudden change in longitude of magnetic field vector. This region lasted for less than an hour, but represents a separate structure between R1 and R2 which could be a magnetic reconnection signature between field lines of region R1 and R2 \citep{Wang2003,Gosling2005}, however an in depth analysis is required to confirm this. In situ observations also reveal that region R2 is overheated $\approx$ 10$^{6}$ K which is because it is squeezed between the region R1 and R3. Region R2 shows a high speed of 750 km s$^{-1}$ at the front while very low speed of 450 km s$^{-1}$ at its trailing edge. Such observations may indicate an extremely fast expansion of R2 due to magnetic reconnection at its front edge as suggested by \citet{Maricic2014}. From an overall inspection of in situ data, it is clear that in situ measured plasma is heated ($\approx$ 10$^{5}$ K for region R1 and R3 and $\approx$ 10$^{6}$ K for region R2 ) than what is observed ($\approx$ 10$^{4}$ K) in general, in CMEs. Such signatures of compression and heating due to CME-CME interaction and passage of CME driven shock through the preceding CME have also been reported in earlier studies \citep{Lugaz2005,Liu2012,Temmer2012,Mishra2014a}. From the in situ data, it is also noted that the spatial scale of CME1 (R1) and CME2 (R2) is smaller than CME3 (R3) and it may be possible due to their compression by the following CME or shock for each.

\subsection{Estimation of Arrival Time of CMEs}
\label{Arrtime}

If the measured 3D speeds (Figure~\ref{CME123_3DFM}) of CME2 and CME3 at the final height is assumed to be constant for the distance beyond COR2 FOV, then CME3 would have caught the CME2 at 39 \textit{R}$_{\odot}$ on 2011 February 15 at 17:00 UT. However, our analysis of HI observations (using J-maps) shows that these two CMEs collided $\approx$ 7 hr earlier (at $\approx$ 28 \textit{R}$_{\odot}$). This can happen due to several reasons. Firstly, because of tracking of two different features in COR and HI observations. Secondly, a deceleration of CME2 beyond COR2 FOV may also be partially responsible for this. Taking 3D speed estimated in COR2 FOV as a constant up to L1, the arrival times of CME1, CME2 and CME3 at L1 will be at 13:00 UT on February 16, at 20:10 UT on February 18 and 23:20 UT on February 17, respectively. However, as discussed in section~\ref{Reconstruction HI}, after the collision between CME2 and CME3, the dynamics of CMEs changed. Therefore, we extrapolated linearly the height-time plot up to L1 by taking few last points in post-collision phase of these CMEs and obtained their arrival time. Such extrapolation may contribute to uncertainties in arrival times of CMEs \citep{Colaninno2013}. From these extrapolations (shown in top panel of Figure~\ref{CME123_SSSE}), the obtained arrival time of CME2 and CME3 at L1 is on 2011 February 18 at 02:00 UT and 05:00 UT, respectively. These extrapolated arrival time for CME2 and CME3 is 12 hr earlier and 6 hr later, respectively, than estimated from measurements made in COR2 FOV. Based on these results, we infer that after the collision of CME2 and CME3, CME2 gained kinetic energy and momentum at the cost of the kinetic energy and momentum of CME3. The arrival time of CME3 is also estimated (within an error of 0.8 to 8.6 hr) by \citet{Colaninno2013} by applying the various fitting approaches to the deprojected height-time data derived by the use of GCS model to SECCHI images.

We associate the starting times of in situ structures marked as R1, R2 and R3 (in Figure~\ref{CME123_insitu}) with the actual arrival of CME1, CME2 and CME3, respectively. We find that marked leading edge of CME1 at L1 is $\approx$ 14 hr earlier than that estimated by extrapolation. The extrapolated arrival time for CME1 is 18:40 UT on February 18. This difference can be explained by assuming a possible acceleration of CME1 beyond the tracked points in HI FOV. We have extrapolated CME1 height-time tracks from its pre-interaction phase because CME1 could not be tracked in \textit{J}-maps up to longer elongations where the interaction is inferred. This highlights the possibility that after its interaction with CME2 or CME3 driven shock (discussed in section~\ref{Reconstruction HI}), the CME1 has accelerated.

The actual arrival time of CME2 and CME3 leading edge (shown in Figure~\ref{CME123_insitu}) is $\approx$ 8.5 and 15 hr later, respectively, than obtained by direct linear extrapolation of height-time curve (Figure~\ref{CME123_SSSE}). From the aforementioned arrival time estimates, we notice an improvement in arrival time estimation of CME2 and CME3 by few (up to 10) hr, when the post-collision speeds are used rather than their speeds before the collision. The average measured (actual) transit speed of CME2 and CME3 at L1 is approximately 100 km s$^{-1}$ larger than its speed in remote observations in post-collision phase. Such an inconsistency of delayed arrival even having larger speeds is possible, only if it is assumed that CME2 and CME3 over-expands before reaching L1 or in situ spacecraft is not hit by the nose of these CMEs \citep{Maricic2014}. The short duration of CME2 in in situ data with lack of magnetic cloud signature favor for a flank encounter of CME2 at the spacecraft. The late arrival of CME3 may also be due to its higher deceleration than estimated in HI FOV. Such inconsistency may also arise, if the remotely tracked feature is incorrectly identified in the in situ data.

\subsection{Geomagnetic Response of Interacting CMEs}
\label{Geomag}

In the bottom panel of Figure~\ref{CME123_insitu}, longitudinally symmetric disturbance (Sym-H) \citep{Iyemori1990} index is plotted. This index is similar to hourly disturbance storm time (Dst) \citep{Sugiura1964} index but uses 1 minute values recorded from different set of stations and slightly different coordinate system and method to determine base values. The effect of solar wind dynamic pressure can be more clearly seen in Sym-H index than in the hourly Dst index. We observed a sudden increase in Sym-H index up to 30 nT around 01:30 UT on February 18 which is within an hour of the arrival of interplanetary shock. The Sym-H index continued to rise, and around 04:15 UT reached 57 nT. We noticed that the first steep rise in this index marked by the shock is  represented by enhanced magnetic field, speed and density. The second peak in Sym-H is primarily due to a corresponding peak in magnetic field strength and density, however no peak in speed was observed during this time. During passage of region R1, the z- component of IP magnetic field (Bz) began to turn negative at 04:07 UT and remained so up to one hour. During this period, its values reached down to -25 nT at 04:15 UT and then turned to positive values around 05:00 UT. We noticed that  Bz turned negative second time at 07:07 UT and remained so for 47 minutes reaching a value of -15 nT at 07:31 UT on February 18. From the Sym-H plot, it is clear that these negative turning of Bz twice, caused a fast decrease in elevated sym-H values. \citet{Dungey1961} has shown that the negative Bz values and process of magnetic reconnection at magnetosphere enables magnetized plasma to transfer its energy into the magnetosphere and form ring current.

Succinctly, we infer that the arrival of magnetized plasma can be attributed to the strong storm sudden commencement (SSC) (Dst = 57 nT) and short duration (47 minutes) negative Bz field therein resulted in a minor geomagnetic storm (Dst = -32 nT). It seems that intensity of SSC is independent of the peak value of depression in the horizontal component of magnetic field during the main phase of a geomagnetic storm. Our analysis supports the idea of collision (or interaction) of multiple CMEs which can enhance the magnetic field strength, density and temperature within CMEs \citep{Liu2012,Mostl2012}. Such enhanced parameters can increase the conductivity of CME plasma, and result in intense induced electric current in CME when it propagate towards the Earth's magnetic field. This induced electric current within CME plasma causes its intense shielding from Earth's field and increases the magnetic field intensity around the Earth which is manifested as SSC \citep{Chapman1931}.

\section{Results and Discussions} 
\label{ResDis} 
In what follows, we summarize our results on the analysis of interaction of three Earth-directed CMEs launched in succession during 2011 February 13-15, on three main aspects. These conclude the morphological study, the kinematic study of interacting CMEs, and then near-Earth manifestations.

\subsection{Morphological Evolution of CMEs}             
 \label{ResDis1}

We have studied the morphological properties of Earth-directed CMEs (CME2 and CME3) when separation angle between \textit{STEREO-A} and \textit{STEREO-B} is 180$^\circ$. On comparing the morphological evolution of CMEs with cone model, we found that slow speed CME2 maintained a constant angular width in the corona, but the angular width of fast speed CME3 decreased monotonically as it propagated further in the corona. The possible explanation for this is that when CME3 is launched from the Sun, its leading edge suddenly experiences the ambient solar wind pressure and result in its flattening \citep{Odstrcil2005}, causing a large angular width. However, as the CME3 propagates further in the corona, there is a decrease in interaction between solar wind and the part of CME (i.e. near apex of the cone) which decides the angular width, therefore, a decrease in angular width is noticed away from the Sun.

The difference in the cone and contour area of CME2 increases linearly with radial height of CME leading edge (bottom panel of Figure~\ref{CME2_area}). This can be explained by the fact that the CME2 interacted with the solar wind such that its leading edge (specially the nose) stretched out  thereby increasing the value of $r$ (distance between Sun-center and nose of CME) and also the ice-cream cone area. For the fast speed CME3, we find that contour area is less than the cone area close to the Sun but as the CME propagated further in corona, its contour area became larger than the cone area (Figure~\ref{CME3_area}). This can be possibly explained by the concept that, contrary to behavior of CME2, as the CME3 propagated further in the corona its front flattened due to drag force, leading to spilling some CME mass outside the cone, i.e. at the flanks of CME. This flattening resulted in a lower estimated value of $r$ and hence a decrease in the estimated cone area. Therefore, a negative value is obtained for the difference between cone area and contour area (bottom panel of Figure~\ref{CME3_area}). From Figure~\ref{CME2_area} and~\ref{CME3_area} (bottom panels), we can say that slow and fast speed CMEs deviate from the cone model differently.

Our analysis shows that the estimated 2D angular width (converted from 3D) follows the same trend as observed in 2D images (Figure~\ref{theta_CMEs}), but has slightly different (within 5\% for CME3 and 15\% for CME2) value at a certain height. We also emphasize that the GCS model parameters ($\gamma$, $\kappa$ and $\alpha$) are very sensitive \citep{Thernisien2009} and can only be fitted with limited accuracy, especially for fast CME whose front gets distorted (see, Figure~\ref{CME3_area}) due to possible interaction with solar wind. Also, the estimation of these parameters depends on the visual agreement between GCS modeled CME and the observed CME, and is dependent on the user. It is to be noted that the minor error in these sensitive parameters can lead to significant errors in 3D edge-on and face-on width of GCS modeled CME. This is the reason, despite a reasonably good agreement between GCS model parameters derived in our study with those derived in \citet{Temmer2014}, the 3D values of angular widths for CME3 do not match well with their results.  Although, we acknowledge that measurements of observed 2D width (Figure~\ref{theta_CMEs}) also has some error (within 5$\arcdeg$), which is quite small than the involved uncertainties in 3D or 2D angular width estimated from GCS model. In light of aforementioned uncertainties and results, further work needs to be carried out to investigate the change in angular width of fast CMEs.

\subsection{Kinematic Evolution of Interacting CMEs}
\label{ResDis2}

The 3D speed and direction estimated for three selected CMEs in COR2 FOV suggest their possible interaction in the IP medium. We have found that CME3 is the fastest among all the three CMEs and shows strong deceleration in the COR2 FOV because of the preceding CME2 which acts as barrier for it. From the analysis of kinematics of CMEs in the heliosphere using stereoscopic methods, we have noted that a collision between CME3 and CME2 took place around 24 \textit{R}$_\odot$ - 28 \textit{R}$_\odot$. As the CME1 was faint and could not be tracked up to HI2 FOV in \textit{J}-maps, we inferred based on the extrapolation of distances that CME2 caught up with CME1 between 138 \textit{R}$_\odot$ to 157 \textit{R}$_\odot$.

It may be noted that using three stereoscopic methods in our study, the estimates of velocity and location of collision for three selected CMEs are approximately same (within a reasonable error of few tens of km s$^{-1}$ and within a few solar radii) to those obtained by \citet{Maricic2014} using single spacecraft method. However, it is worth to investigate the relative importance of using single HI observations and simultaneous HI observations from twin STEREO viewpoints for several CMEs launched in different directions at different STEREO spacecraft separation angle.

We have identified signatures of collision between CMEs in the kinematics profiles as exchange in their speed. We analysed momentum and energy exchange during collision phase of CME2 and CME3 and found that the nature of collision was in inelastic regime, reaching close to elastic. This is in contrast to the finding of \citet{Shen2012} who have reported a case of interacting CMEs in super-elastic regime. Also, in another study we have found a case of collision of CMEs close to perfectly inelastic in nature \citep{Mishra2014a}. Therefore, it is worth to investigate further what decides the nature of collision and which process is responsible for magnetic and thermal energy conversion to kinetic energy to make a collision super elastic. Further in-depth study is required to examine the role of duration of collision phase and impact velocity of CMEs for deciding the nature of collision. 

The time variations of estimated direction of propagation of CMEs (Figure~\ref{CME123_SSSE}) shows a surprisingly large change towards the sunward (entrance) edge of HI1 FOV. As explained in section~\ref{CompKinem} these variations are not physical (real) and are mainly due to uncertainties in the measurements of elongation angles. We emphasize that in our analysis of collision dynamics of CMEs, we do not expect large errors to switch in our results because the speed (derived from the distance) has smaller errors (as shown in Figure~\ref{CME123_SSSE}) in comparison to the direction. Moreover, the estimated uncertainties in the derived speed and direction are relatively smaller during the collision phase of the CMEs. Also, we have considered the sufficiently larger uncertainties in the mass and speed, to estimate the nature of collision. Therefore, we advocate that the analysis carried out in this study is reliable.

The present analysis for collision dynamics may have small uncertainties due to adopted boundary for the start and end of collision phase. It is often difficult to define the start of collision as the following CME (CME3) starts to decelerate (due to its interaction with preceding CME) and preceding CME (CME2) start to accelerate before (most possibly due to shock driven by following CME) they actual merge as observed in HI FOV. Also, the assumption that there is no mass transfer between CME2 and CME3 during collision may result in some uncertainties. Further, we have not taken into account the expansion velocity and propagation direction of centroid of CMEs, which may be different before and after the collision. In our study, we found that even after considering reasonable uncertainties in derived mass and velocity parameters, the coefficient of restitution ($e$) lies between 0.78 to 0.90 for the interacting CME2 and CME3. This implies that the total kinetic energy of the system of CMEs after the collision is less than its value before the collision. In our analysis, we used total mass of CMEs to study their collision dynamics, but as the CME is not a solid body therefore its total mass is not expected to participate in the collision. Keeping in mind, various limitations of the present study, we believe that more detailed work, by incorporating various plasma processes, is required to understand the CME-CME interaction.

\subsection{Interacting CMEs Near the Earth}
\label{ResDis3}

We have examined the interaction and the collision signatures of CMEs in the in situ (WIND) observations. The interacting CMEs could be identified as a separate entity in in situ observations, therefore could not be termed as complex ejecta as defined by \citet{Burlaga2002}. The in situ observations suggest that a shock launched by the fastest CME (CME3) passed through the CME2 and CME1 and caused compression, heating and acceleration, in particular for the CME2 which is sandwiched between preceding CME (CME1) and the following CME (CME3). Our analysis shows that the interacting CMEs resulted in a minor geomagnetic storm with a strong long duration SSC. This is in contrast to the results of \citet{Farrugia2004,Farrugia2006,Mishra2014a}, which suggested that interaction of CMEs lead to long duration southward component of magnetic field and therefore to strong geomagnetic storms.

We have found that using the kinematics derived from the stereoscopic methods, in our study, the estimated arrival times of CMEs are slightly (only by few hr) better than those of \citet{Maricic2014}. This does not raise questions on the efficacy of stereoscopic methods and long tracking of CMEs using J-maps, but it simply demonstrates that speeds determined from stereoscopic methods and single spacecraft HM method is approximately same for the selected Earth-directed CMEs during STEREO separation angle of 180$\arcdeg$.

\section{Conclusions}
\label{Conclu}

Based on our analysis of interacting CMEs of 2011 February 13-15 by combining the wide angle imaging and in situ observations, we conclude the following points:

(1) The collision between CME2 and CME3 is observed at 24-28 \textit{R}$_\odot$ while the collision between CME1 and CME2 is inferred at 138-157 \textit{R}$_\odot$. This highlights that heliospheric imaging is important to observe the collision of CMEs and estimate their post-collision dynamics.

(2) We find that the observed collision of CME2 and CME3 is in inelastic regime reaching close to elastic, while earlier studies have shown the nature of collision of CMEs as super elastic \citep{Shen2012} and close to perfectly inelastic \citep{Mishra2014a}. Therefore, further investigations of interacting CMEs are required to understand the nature of collision. 

(3) The total kinetic energy of the CMEs after the observed collision is reduced by 1.3\% of its value before the collision. The exchange of momentum between interacting CMEs range from 35\% to 68\% of its values before the collision. 

(4) The in situ measurements of these CMEs near 1 AU shows that preceding CME1 and CME2 are accelerated, compressed and heated by overtaking CME3 and the shock driven by it. 

(5) Our results do not favor the possibility of strengthening of the geomagnetic response as a consequence of arrival of two or more interacting CMEs at near the Earth. In fact, the interacting CMEs of February 13-15 lead to a minor geomagnetic storm (Dst $\approx$ -32 nT), although a strong long duration SSC (Dst $\approx$ 57 nT) is noticed. This finding is in contrast to earlier inferences by \citet{Farrugia2004,Farrugia2006,Mishra2014a}. 

(6) The morphological evolution of CMEs propagating with slow and fast speed, as compared to ambient solar wind speed, seem to be different. 

Our study of interacting CMEs highlight the importance of Heliospheric Imager (HI) observations and their association with in situ observations to understand the nature of CME-CME interaction in detail and for improved prediction of CME arrival time using post-interaction kinematics. We have also highlighted the difficulties inherent in reliably understanding the kinematics, arrival time, nature of collision as well as morphological evolution of CMEs.

\section*{Acknowledgments}
The authors thank Jackie A. Davies and Manuela Temmer for their valuable comments and useful discussions. We acknowledge the team members of \textit{STEREO} COR and HI as well as WIND instruments. We acknowledge the UK Solar System Data Center for providing the processed Level-2 \textit{STEREO}/HI data. We also express our appreciation to the referee for his/her insightful comments.

\clearpage


\clearpage

\begin{figure}[h]
 \centering
  \includegraphics[scale=0.33]{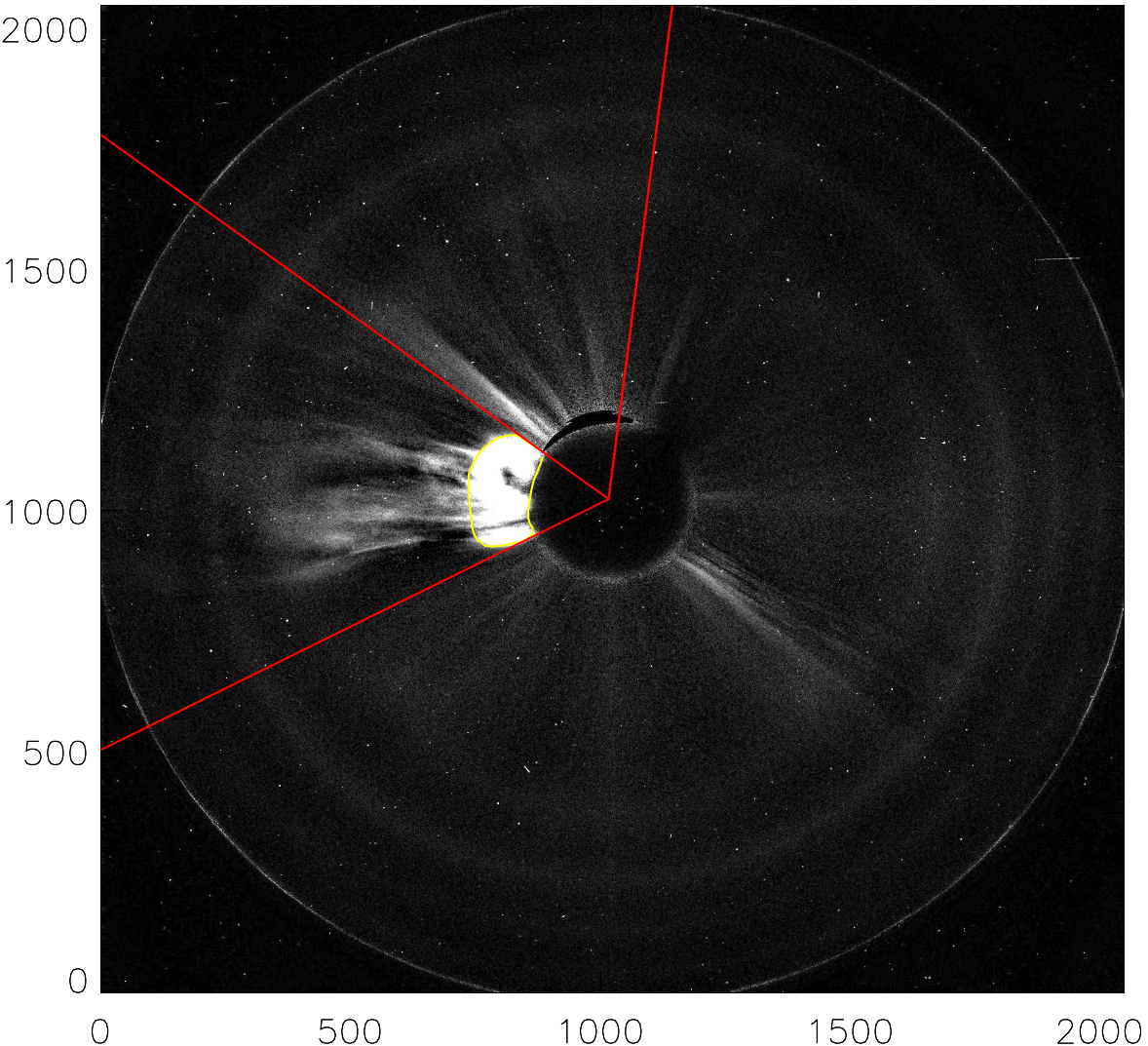}
  \includegraphics[scale=0.33]{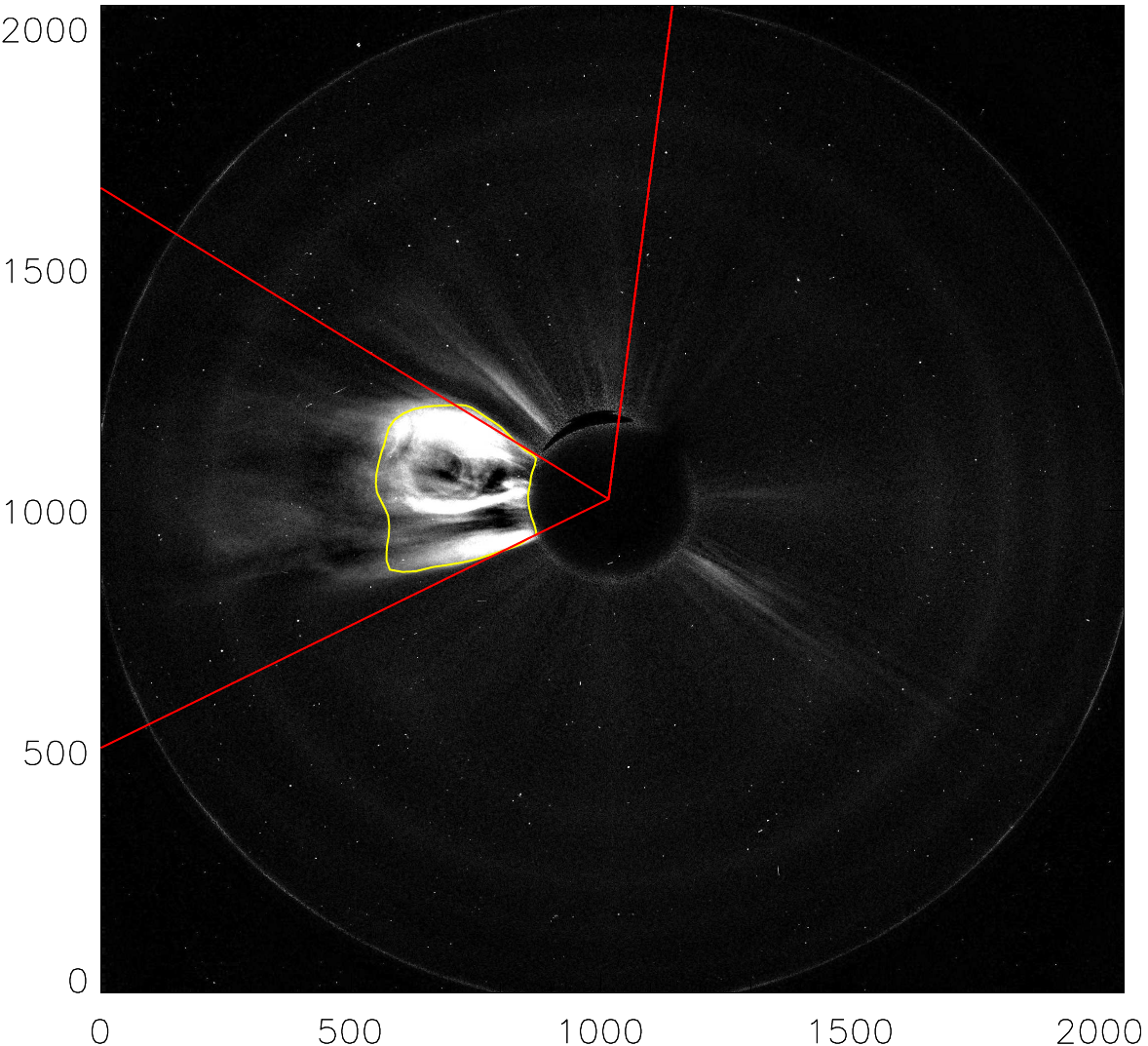}
	\includegraphics[scale=0.33]{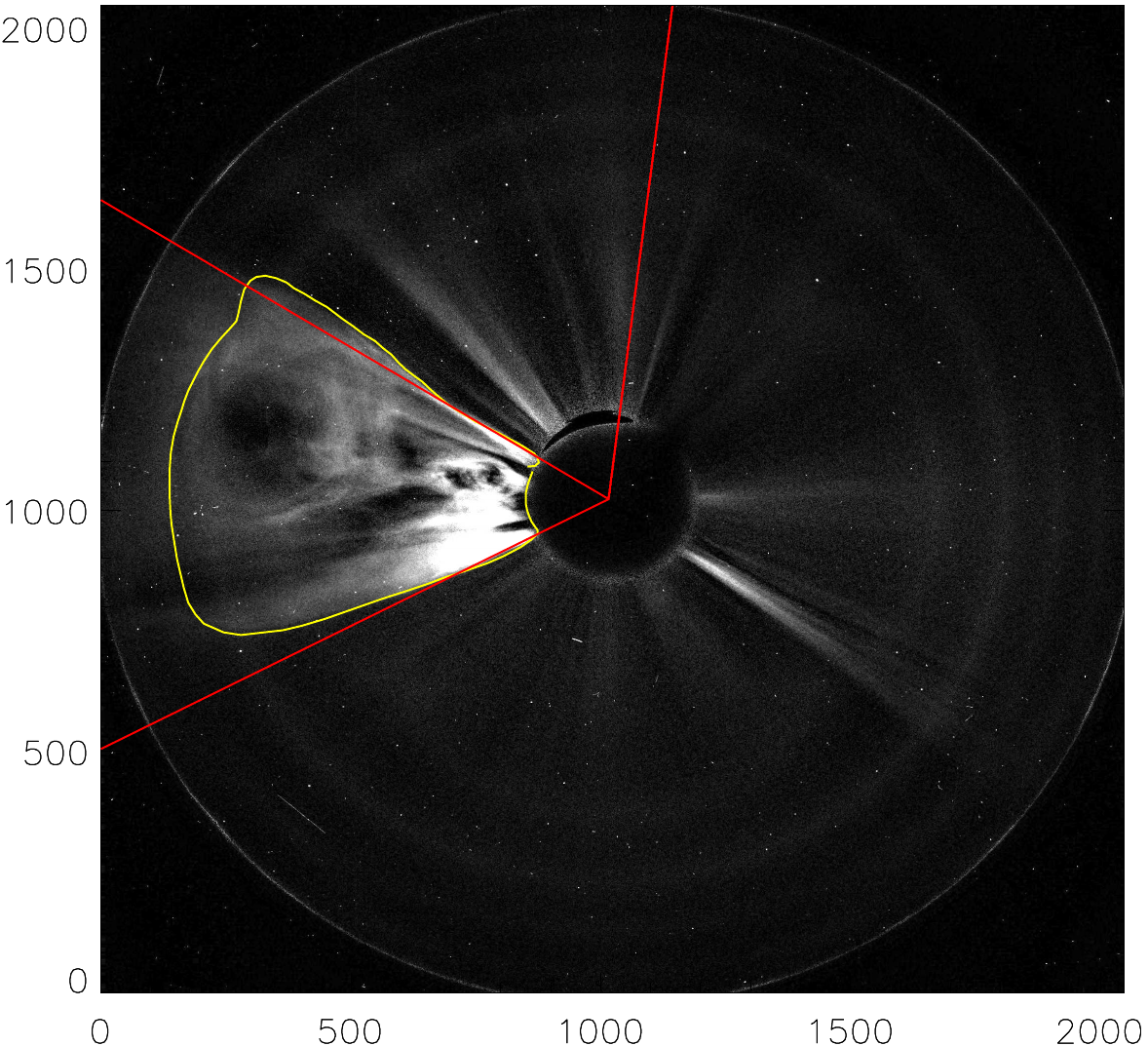}
 
  \includegraphics[scale=0.33]{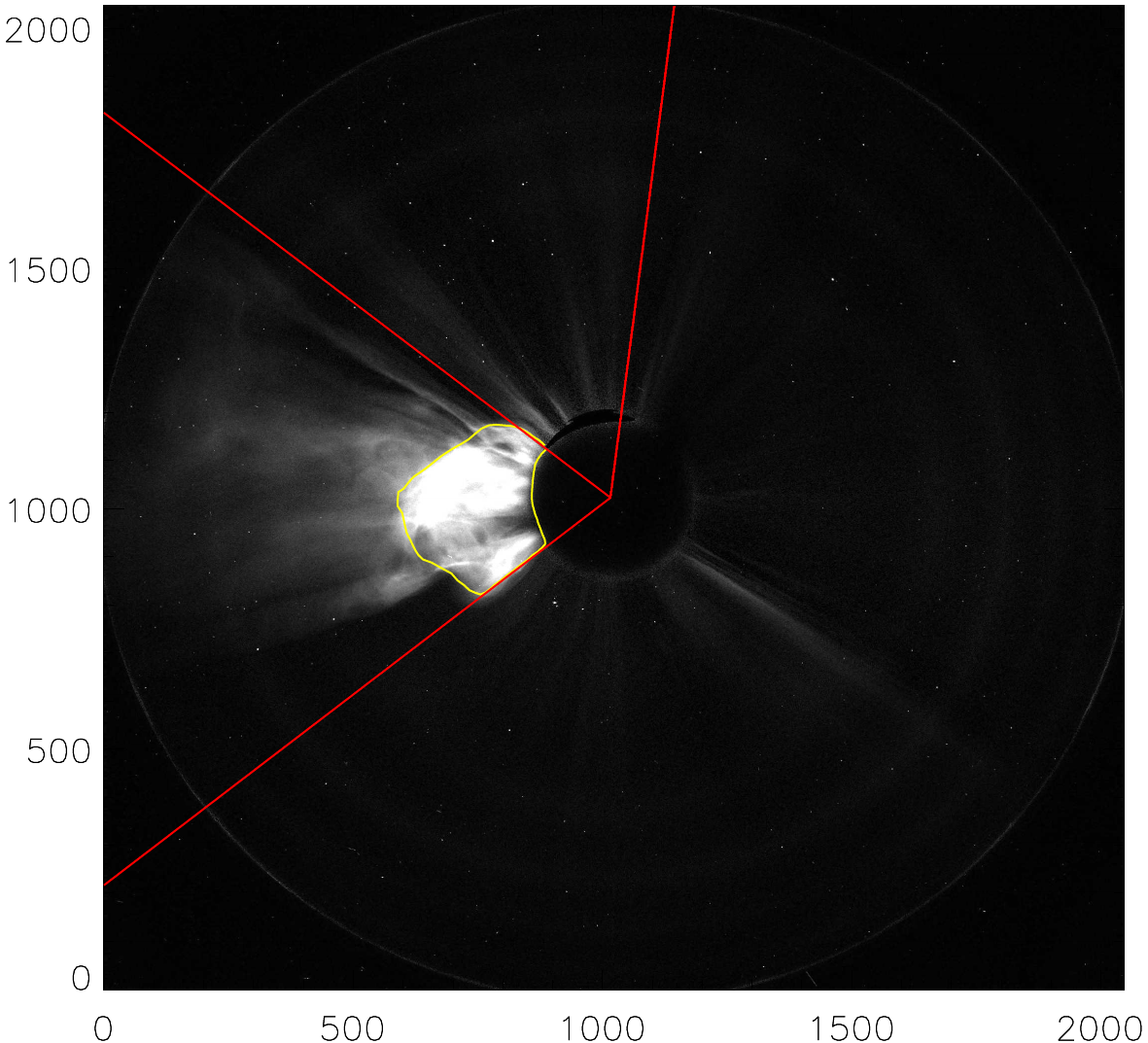}
  \includegraphics[scale=0.33]{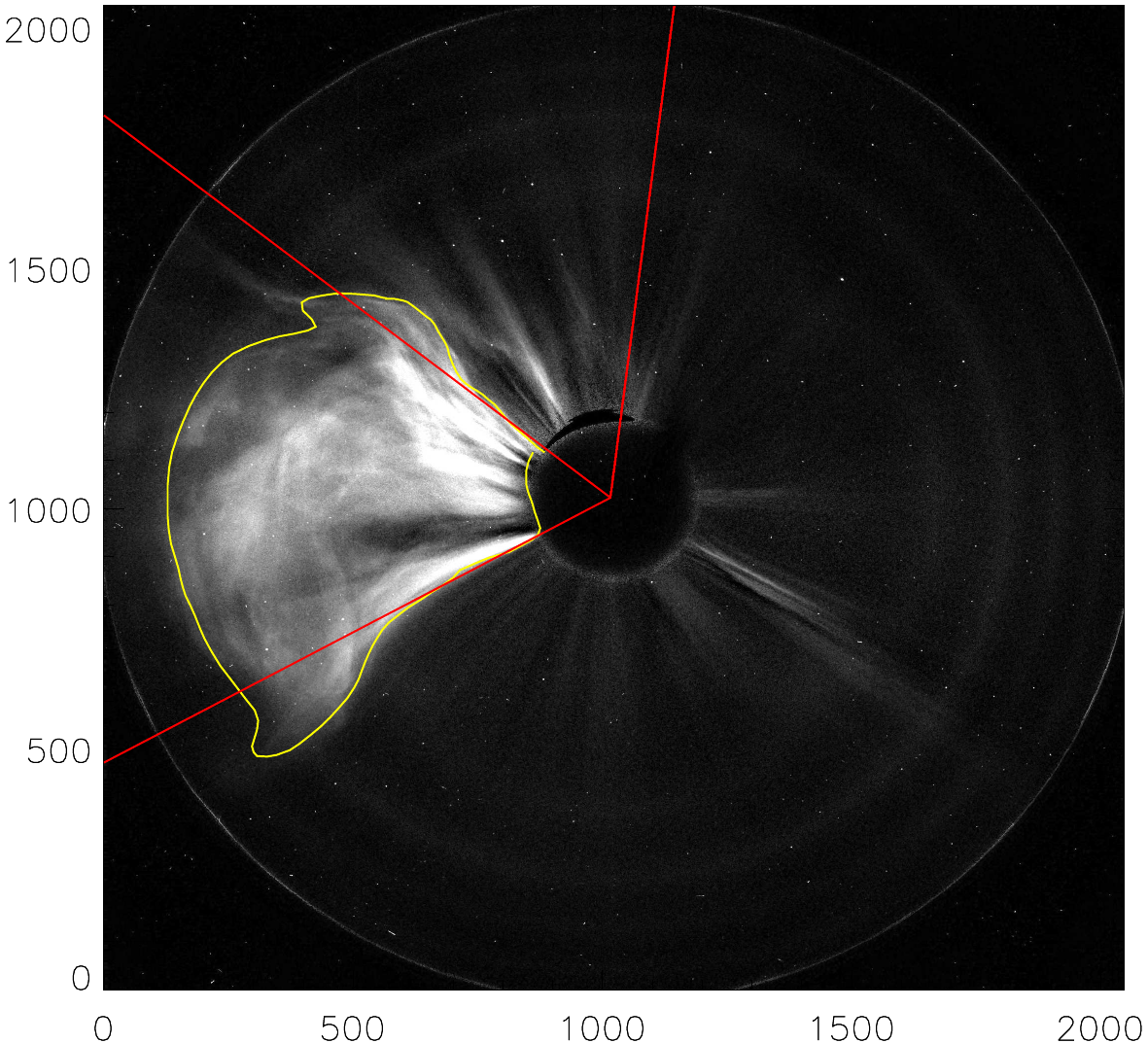}
	\includegraphics[scale=0.33]{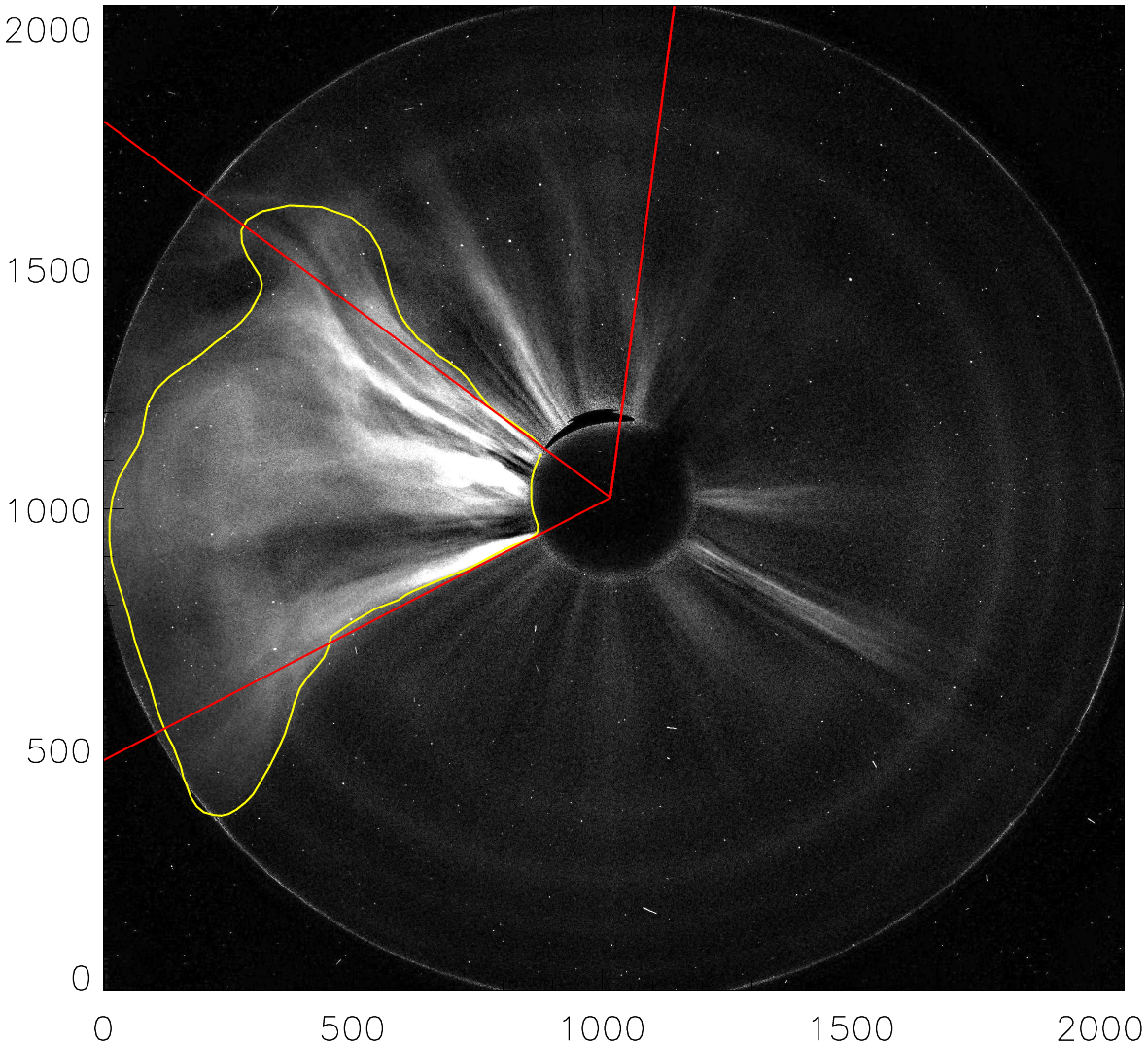}
 \caption{The left, middle and right images in top panel of figure show CME2 in COR2-A FOV at 18:24 UT, 20:24 UT and 23:54 UT on 14 February, respectively. Similarly, left, middle and right images in bottom panel show CME3 in COR2-A FOV at 02:39 UT, 04:24 UT and 05:54 UT on 15 February, respectively. The vertical red lines mark the zero degree position angle in helio-projective radial coordinate system. The other two red lines forming the edges of CME cone are marked at the position angle of CME flanks. The contour with yellow curve encloses the CME area completely.}\label{CME2_contour}
 \end{figure}

\begin{figure}[h]
 \centering
  \includegraphics[scale=0.8]{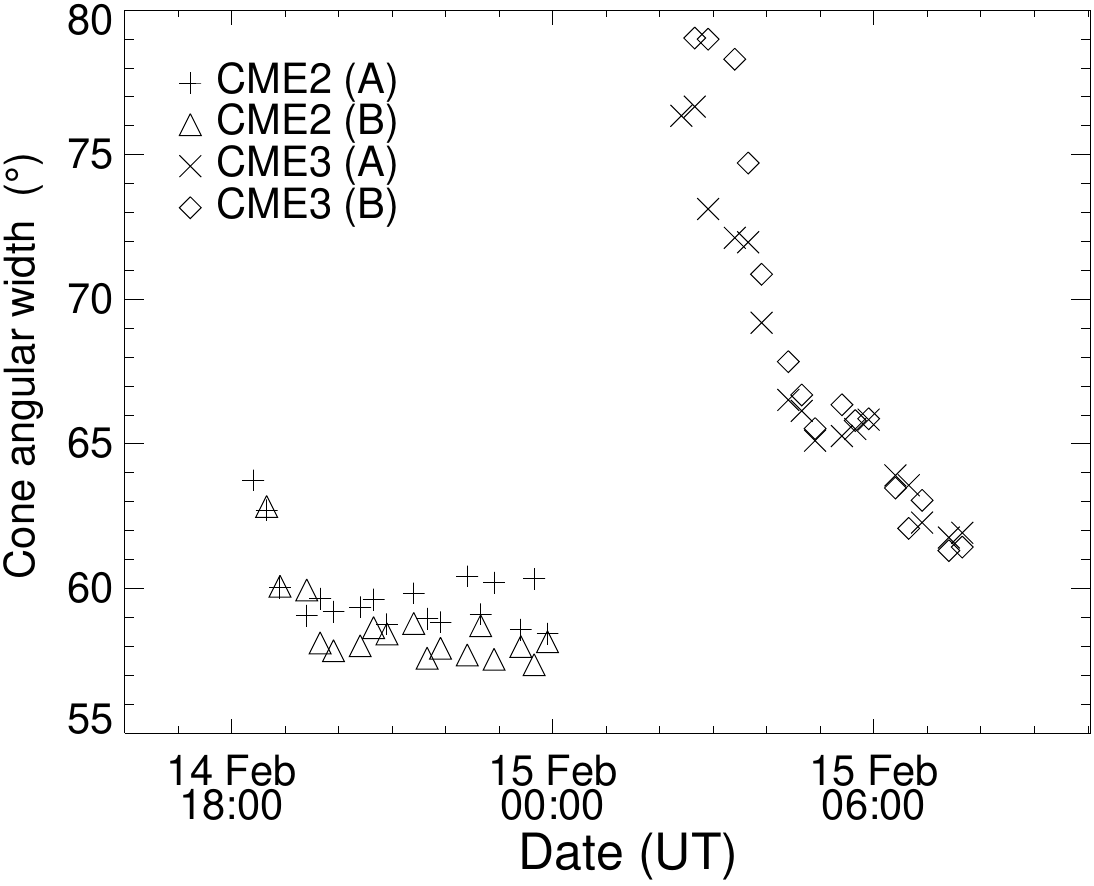}
 \caption{Time variation of estimated 2D cone angular width of CME2 and CME3 from both COR-A and COR-B images.}\label{theta_CMEs}
 \end{figure}

\begin{figure}[h]
 \centering
  \includegraphics[scale=0.8]{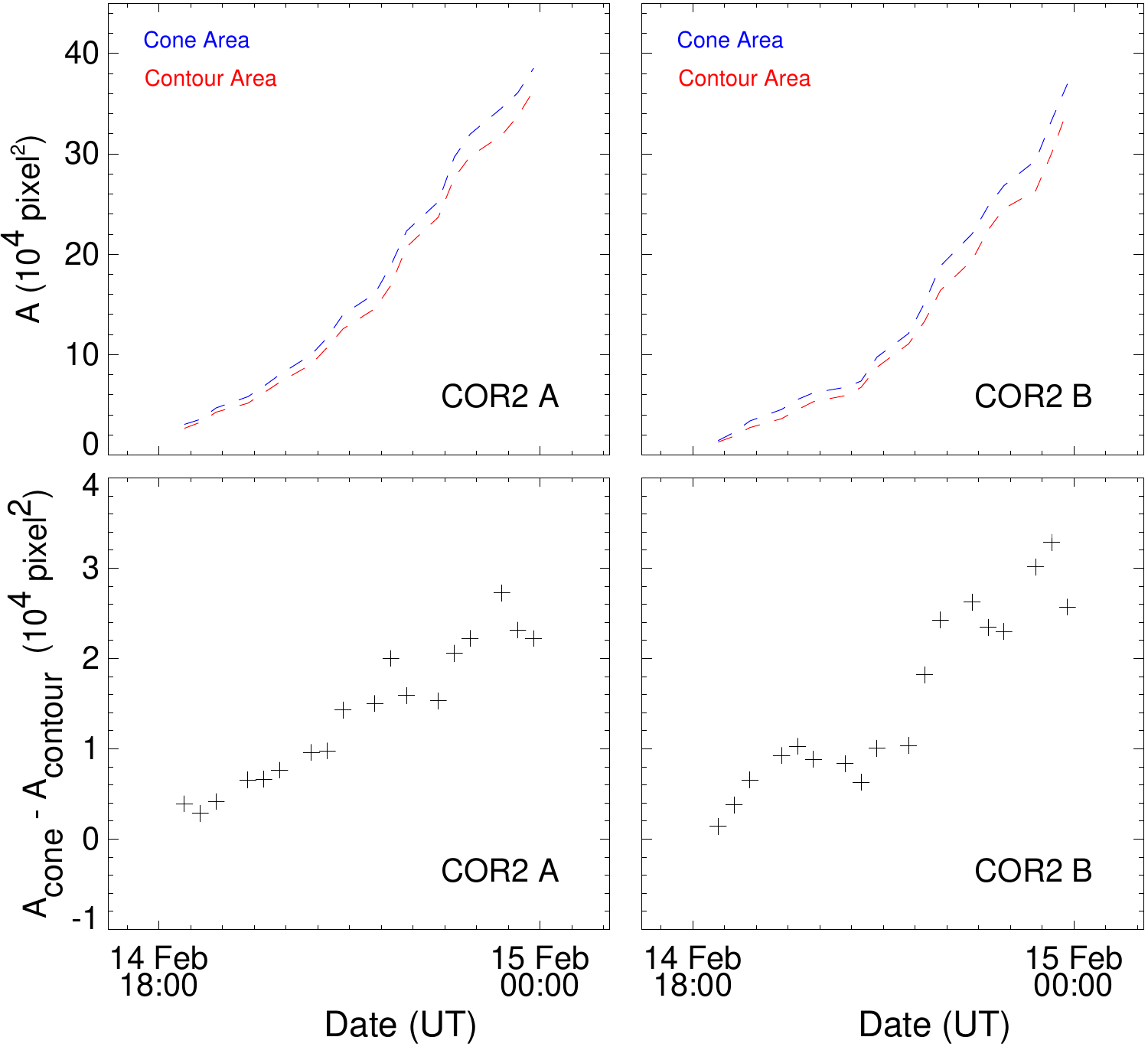}
 \caption{In the top left panel, time variations in cone and contour area of CME2 estimated in COR2-A FOV is shown with blue and red, respectively. Its  variations in COR2-B FOV is shown in top right panel. In the bottom left and right panels, the difference in cone and contour area, estimated for COR2-A and B FOV, respectively, are shown.}\label{CME2_area}
 \end{figure}

\begin{figure}[h] 
 \centering
  \includegraphics[scale=0.8]{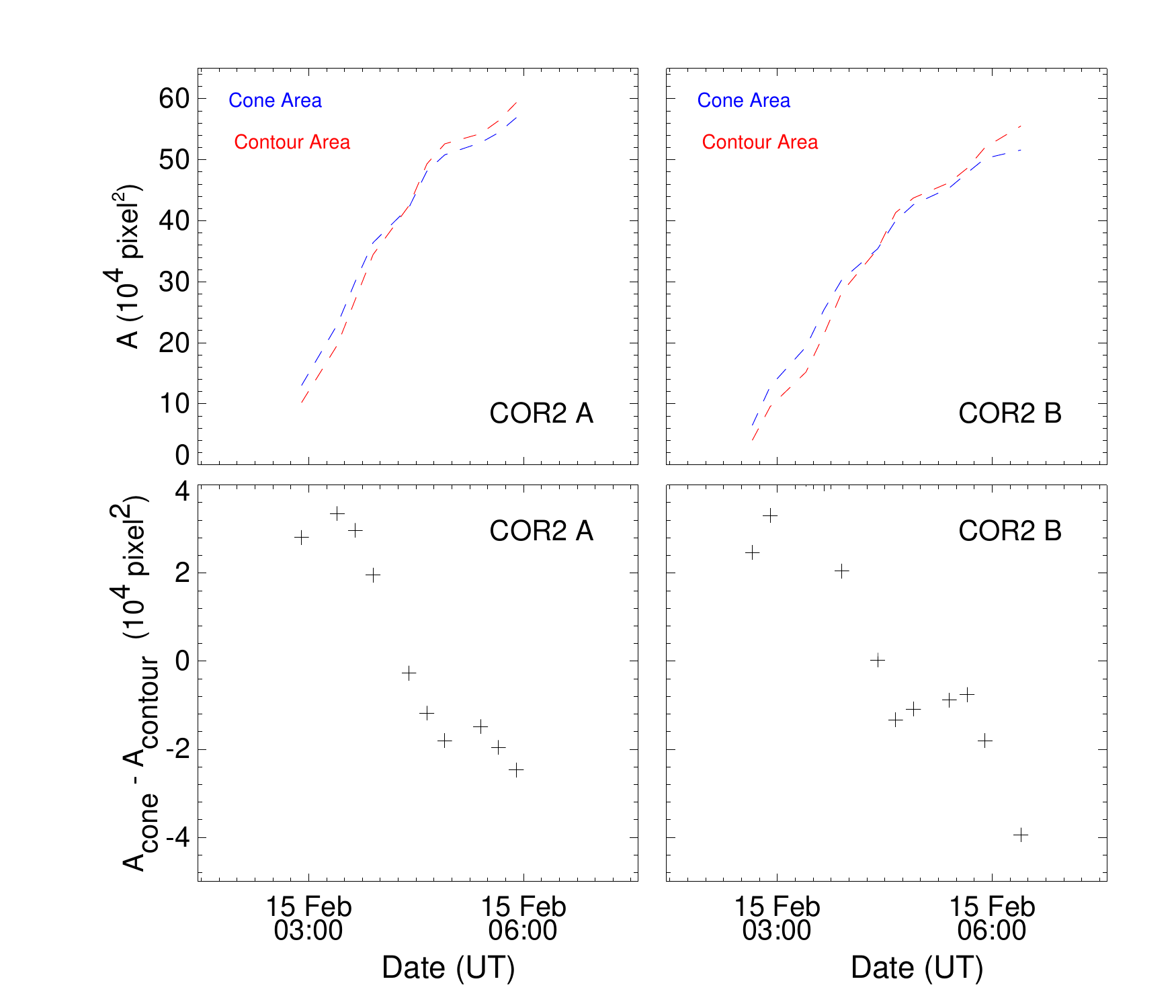}
 \caption{As Figure~\ref{CME2_area}, for the CME3.}\label{CME3_area}
 \end{figure}

\begin{figure}[h]
 \centering
 \includegraphics[scale=0.38]{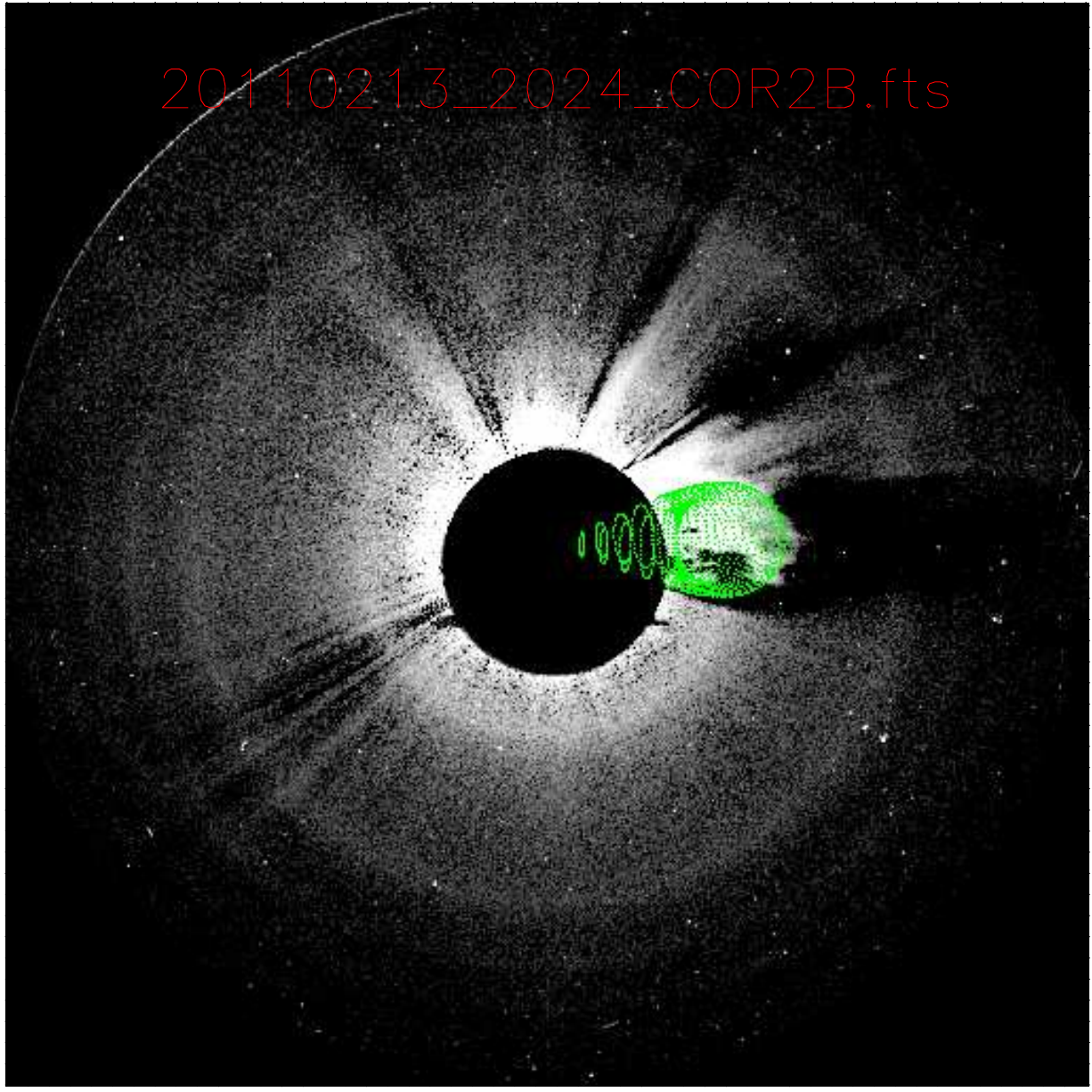}
 \includegraphics[scale=0.38]{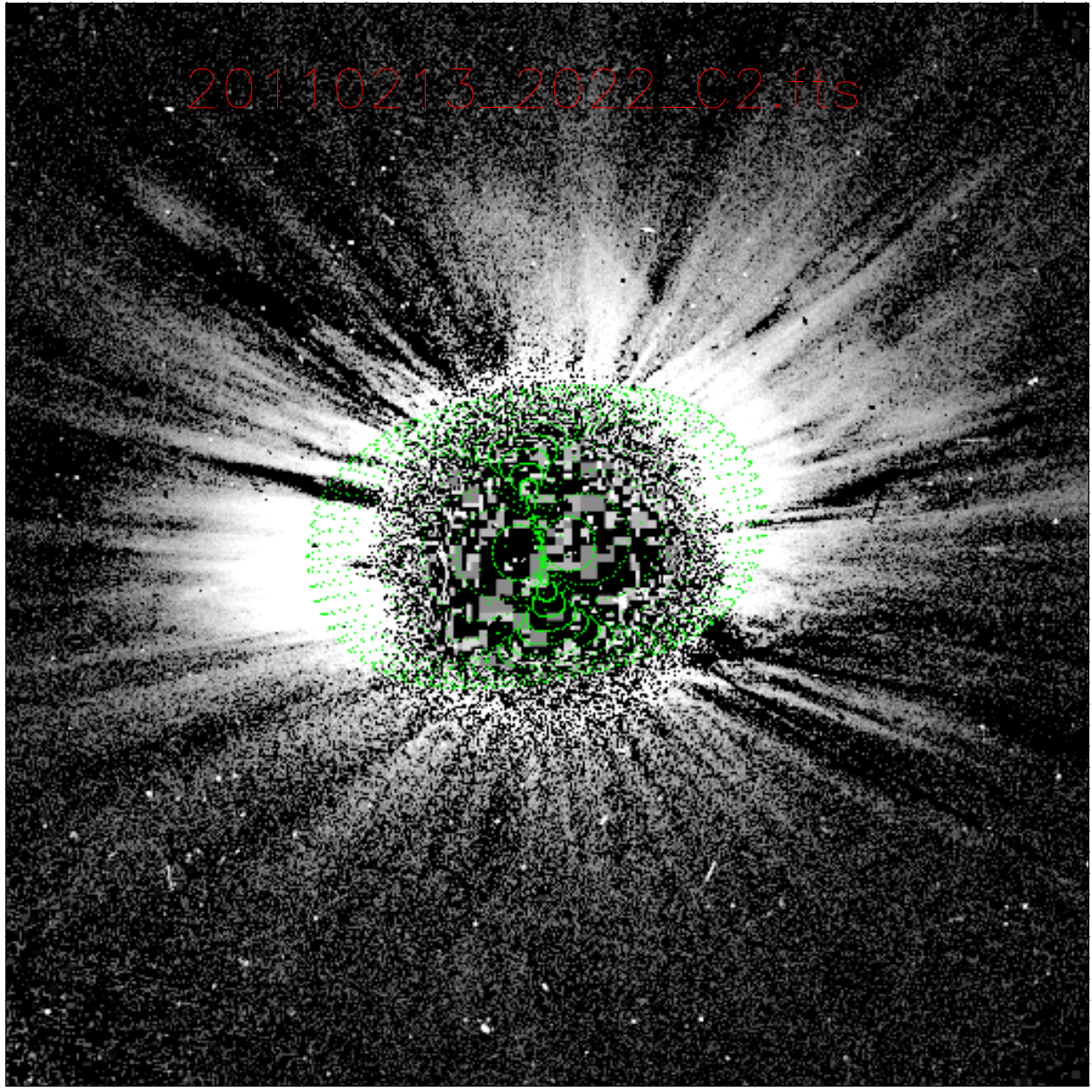}
 \includegraphics[scale=0.38]{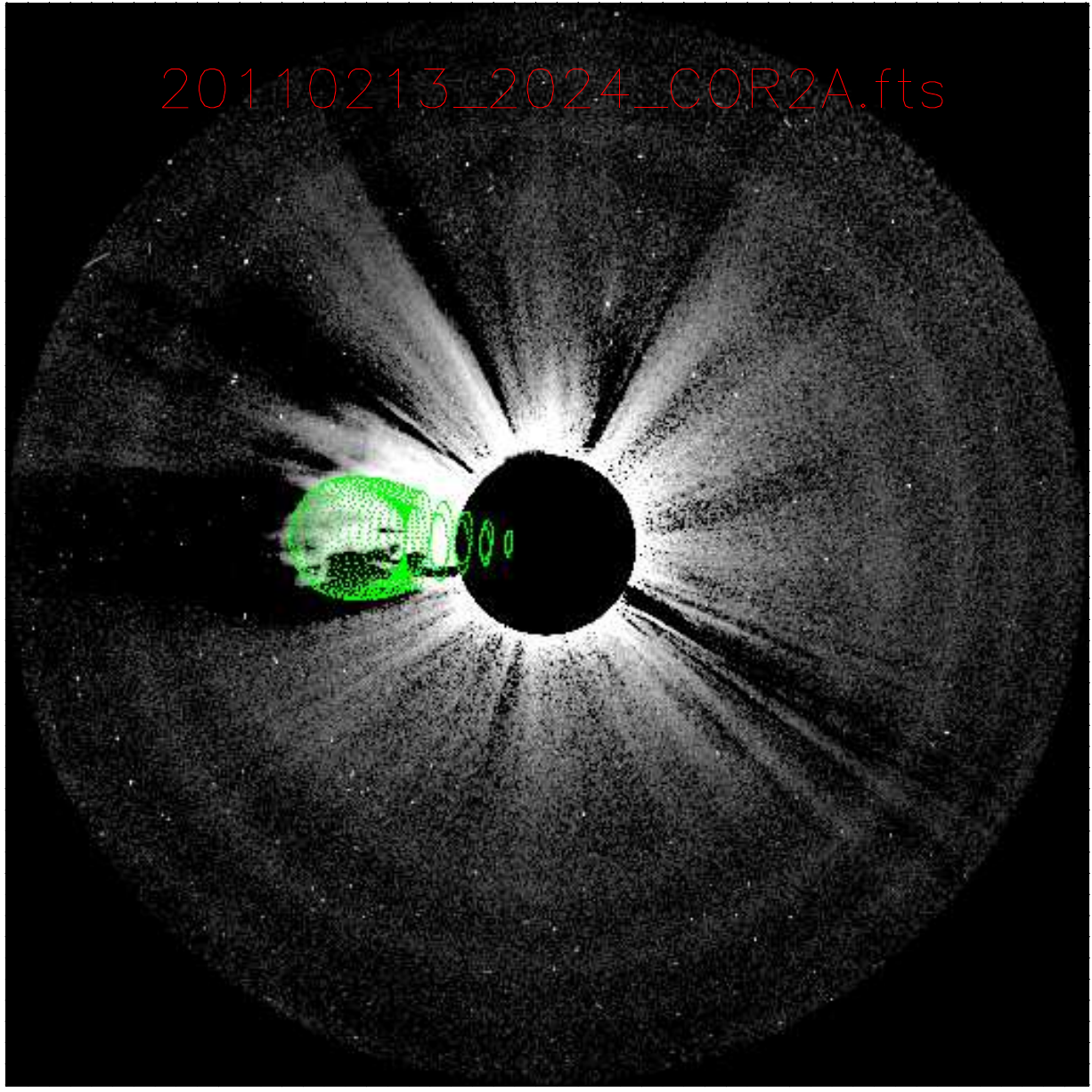}

  \includegraphics[scale=0.38]{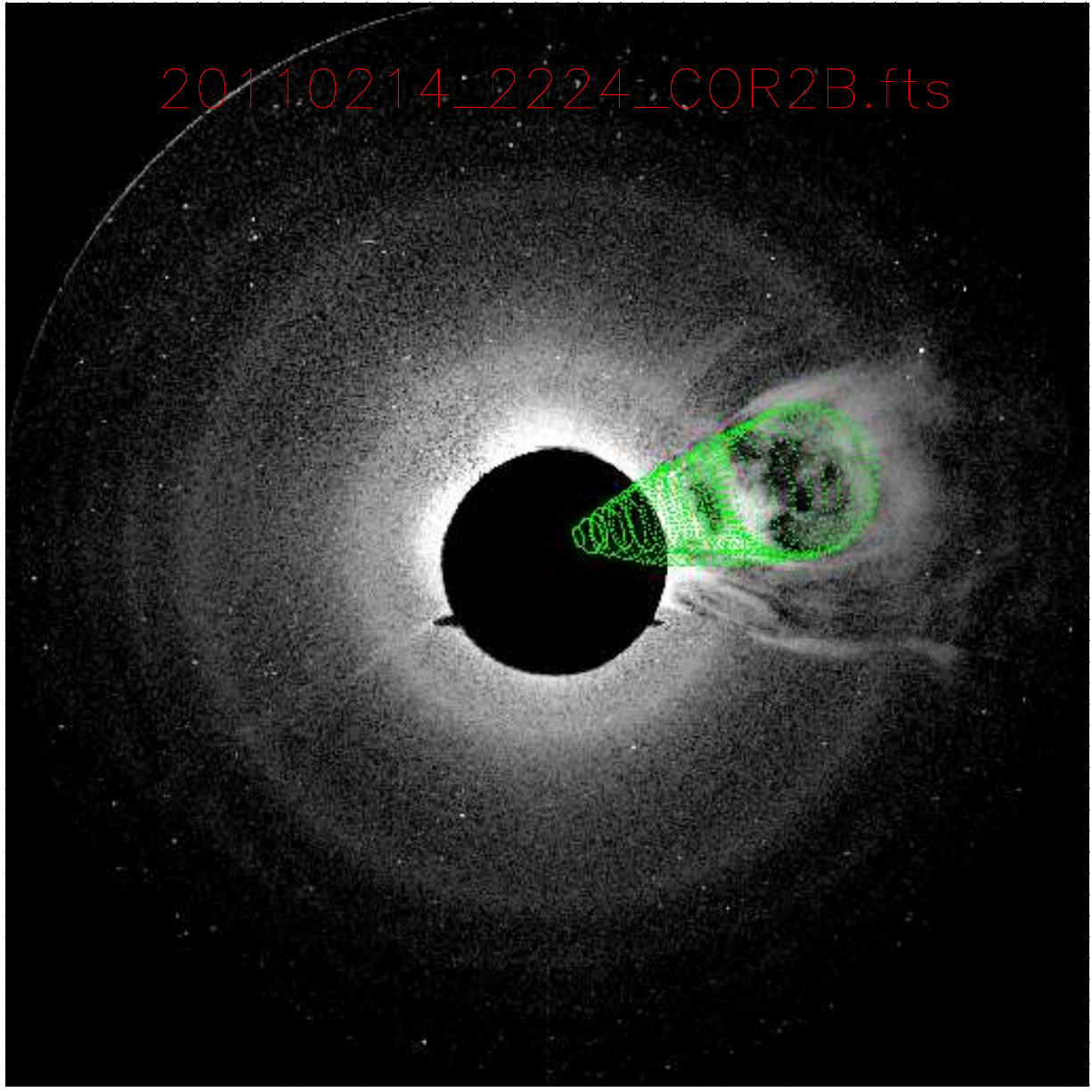}
	\includegraphics[scale=0.38]{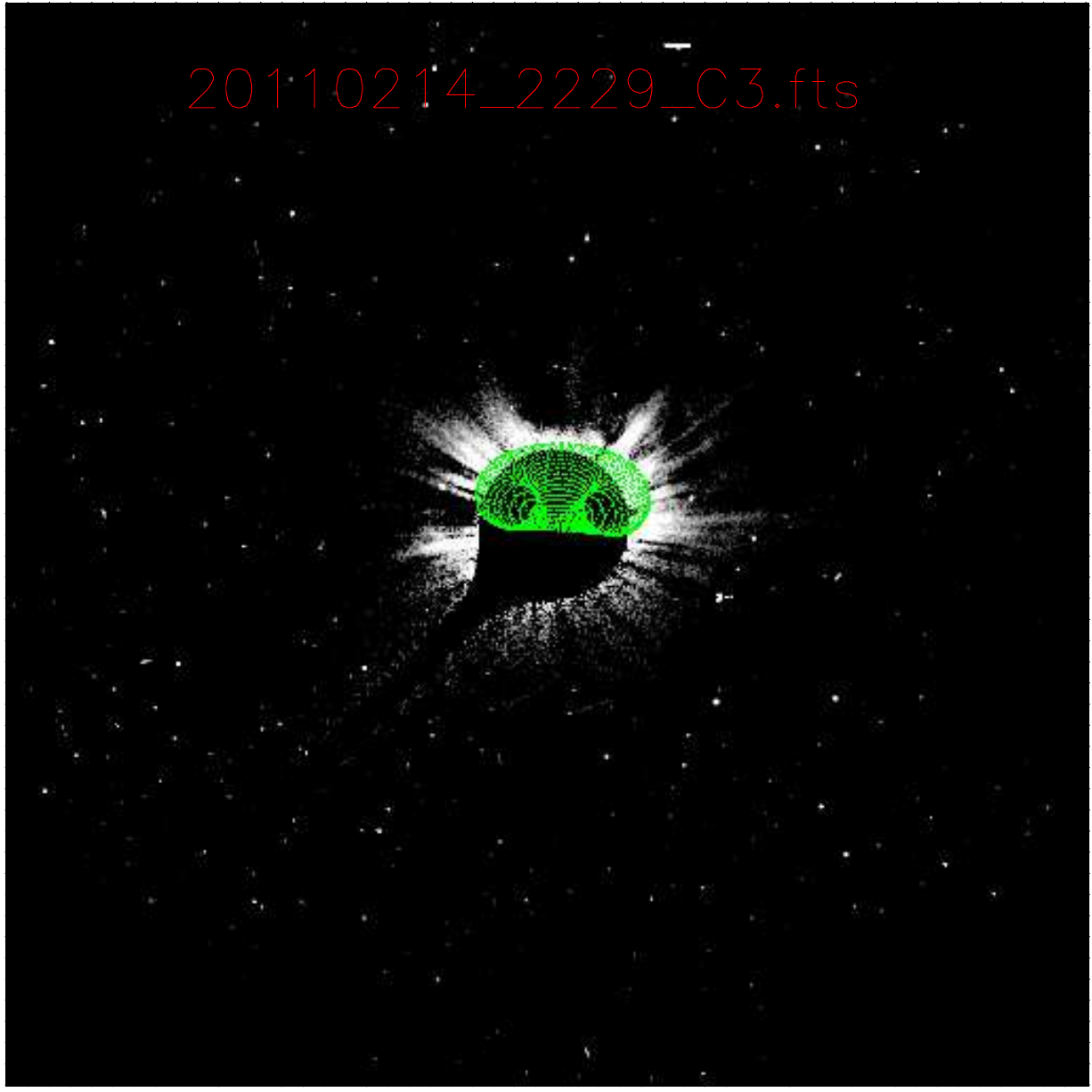}
	\includegraphics[scale=0.38]{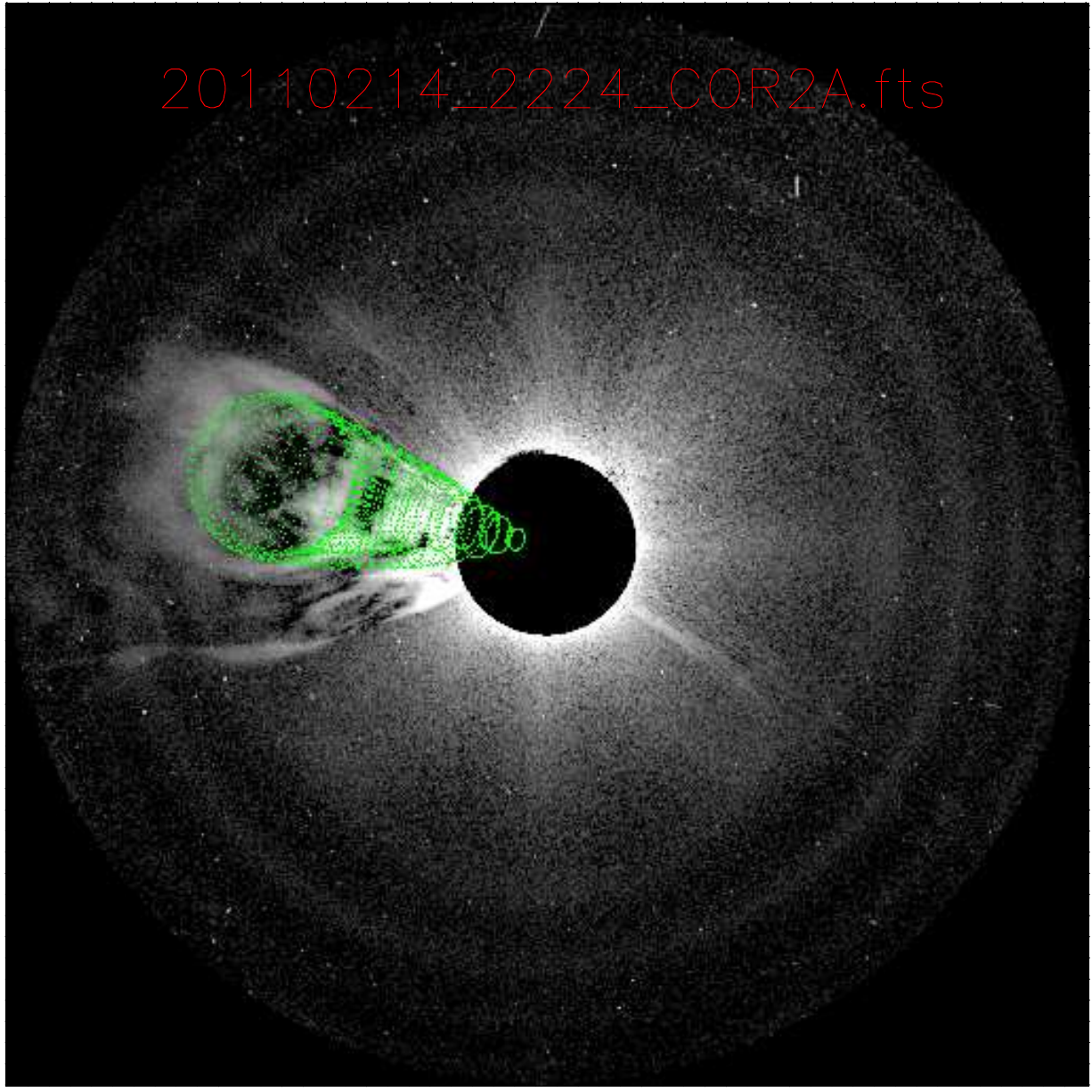}

	\includegraphics[scale=0.38]{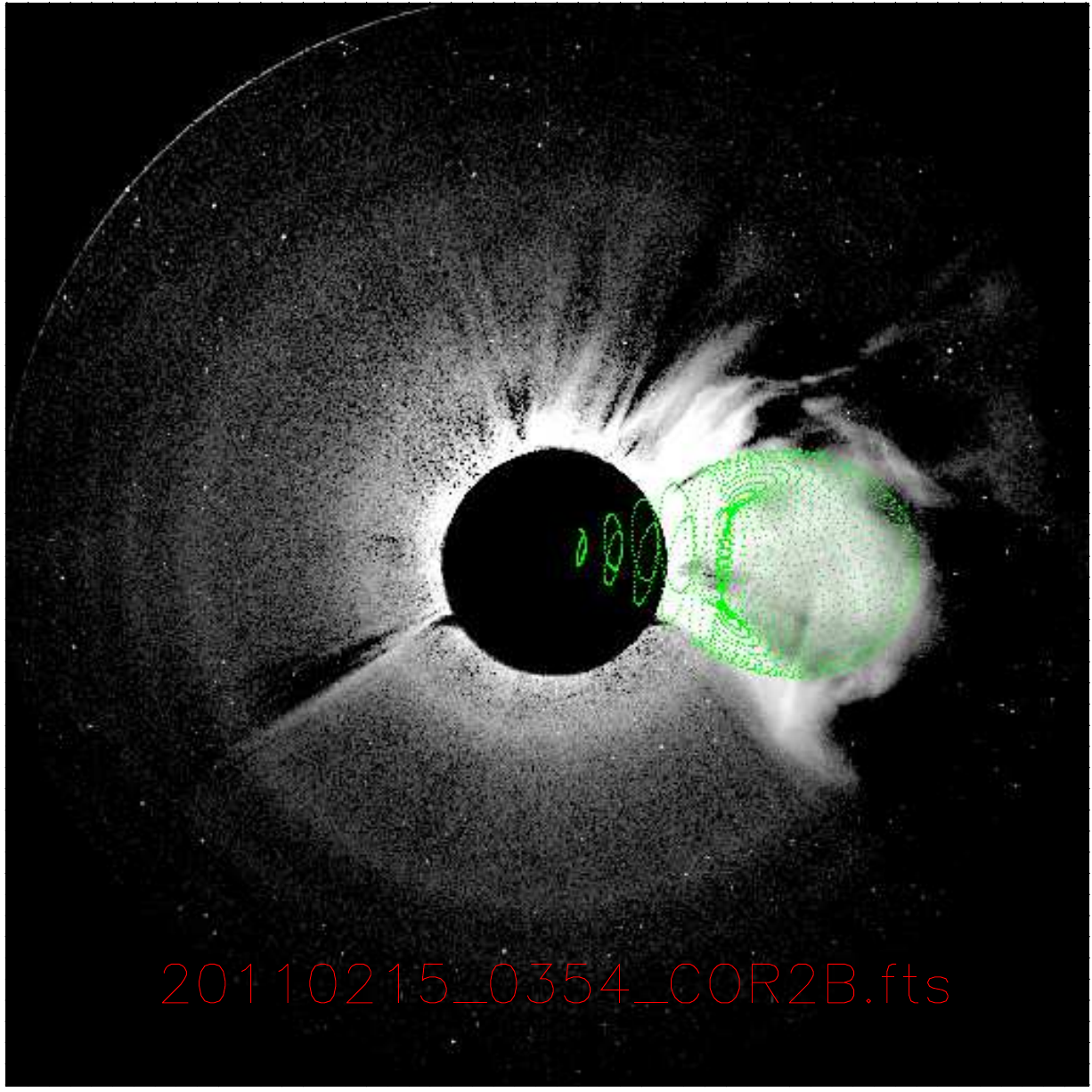}
	 \includegraphics[scale=0.38]{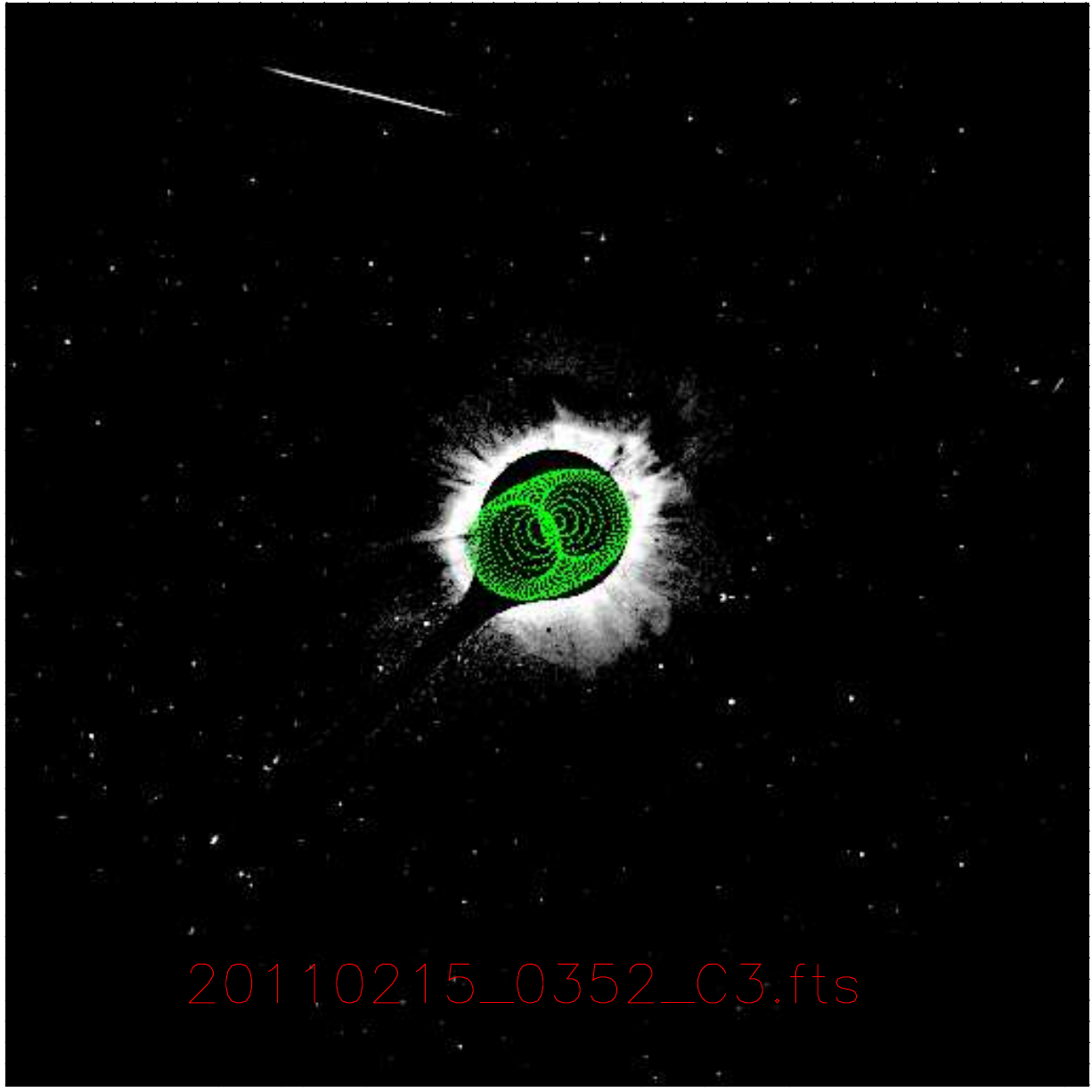}
	 \includegraphics[scale=0.38]{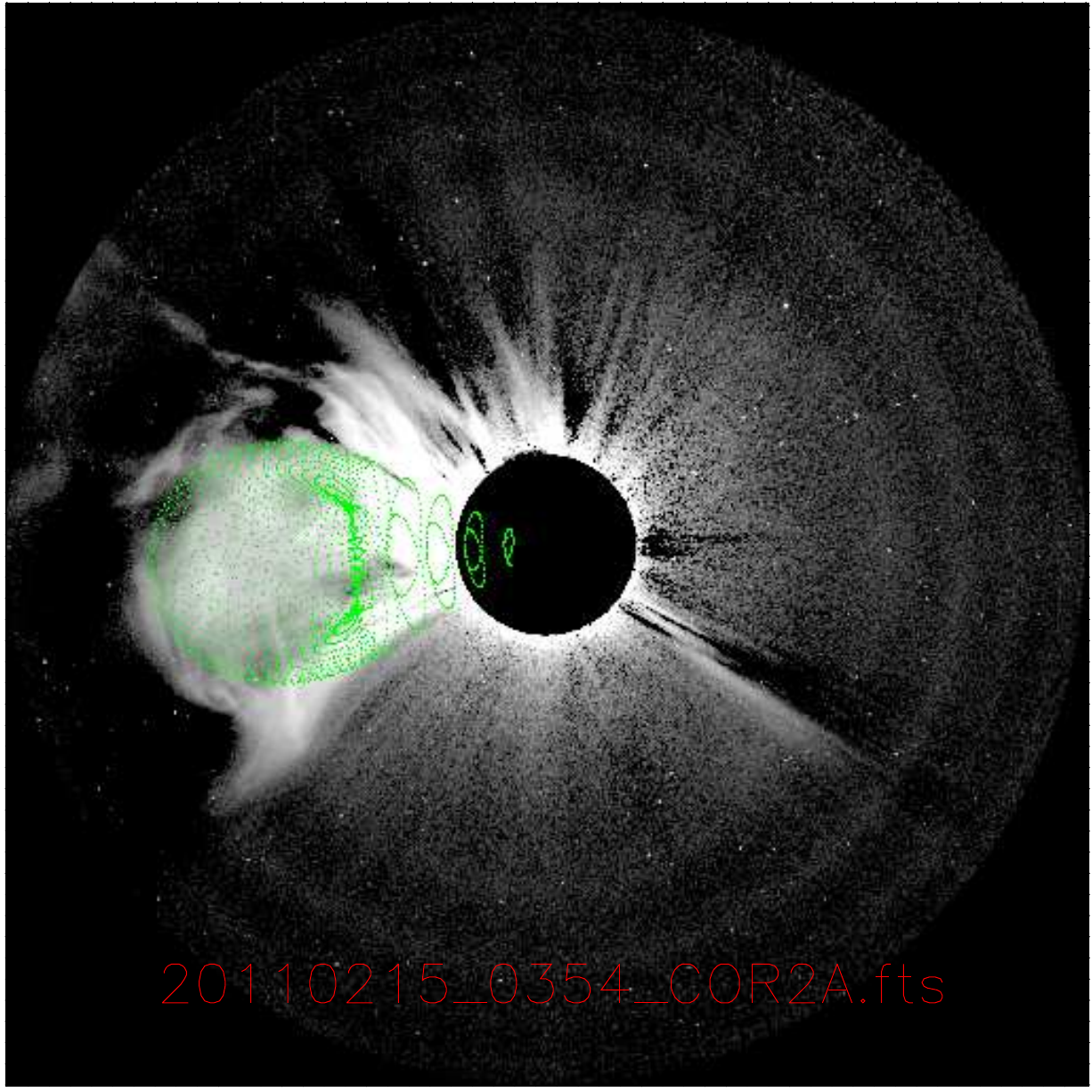}
	
 \caption{The contemporaneous image triplets for CMEs from SECCHI/COR2-B (left), SOHO/LASCO (C2 or C3) (middle) and SECCHI/COR2-A (right) are shown with GCS wireframe (with green) overlaid on it. The top, middle and bottom panels show the images of CME1 around 
20:24 UT on February 13, CME2 around 22:24 UT on February 14 and CME3 around 03:54 UT on February 15, respectively.}\label{CME123_FM}
 \end{figure}

\begin{figure}[h]
 \centering
  \includegraphics[scale=0.95]{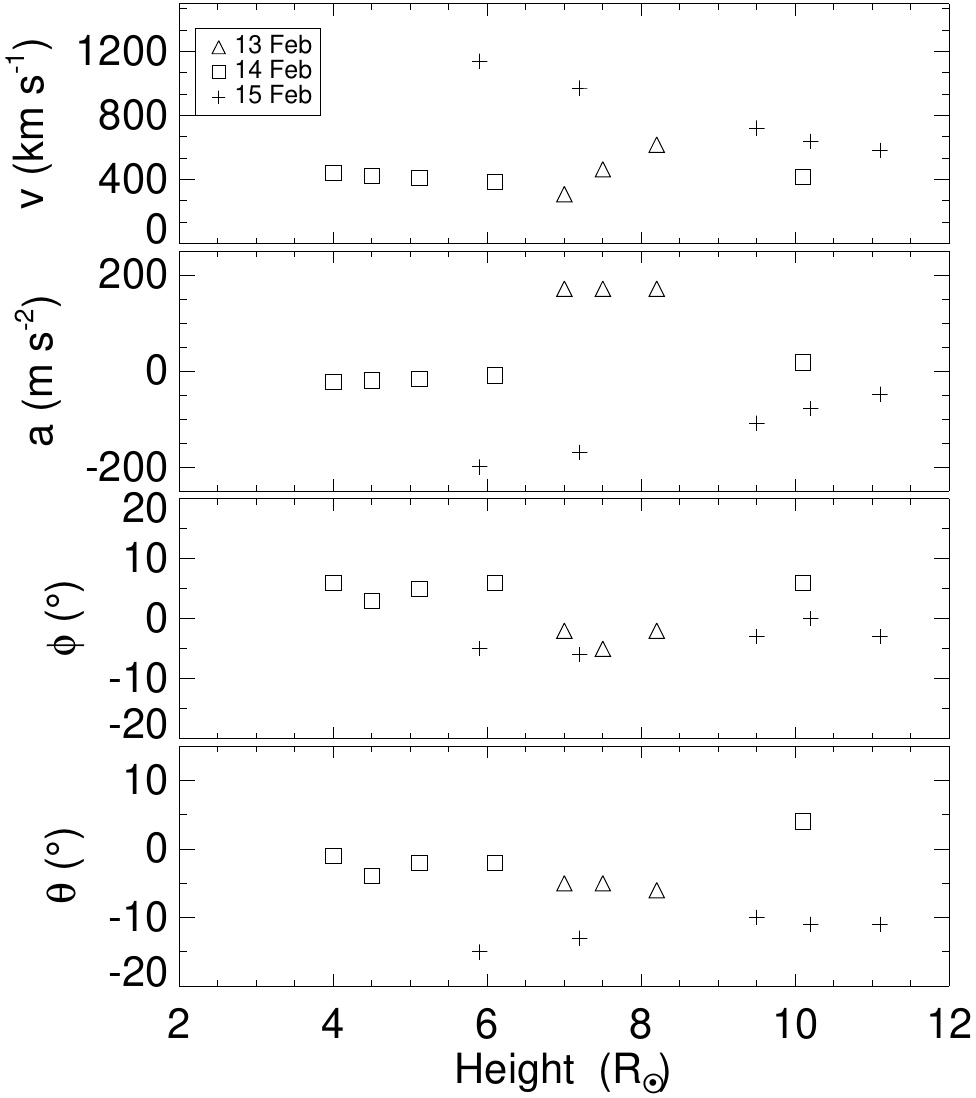}
 \caption{Top to bottom panels show the variations of radial velocity, acceleration, longitude and latitude of CME1, CME2, and CME3 with radial height from the Sun.}\label{CME123_3DFM}
 \end{figure}

\begin{figure}[h]
 \centering
  \includegraphics[scale=0.5]{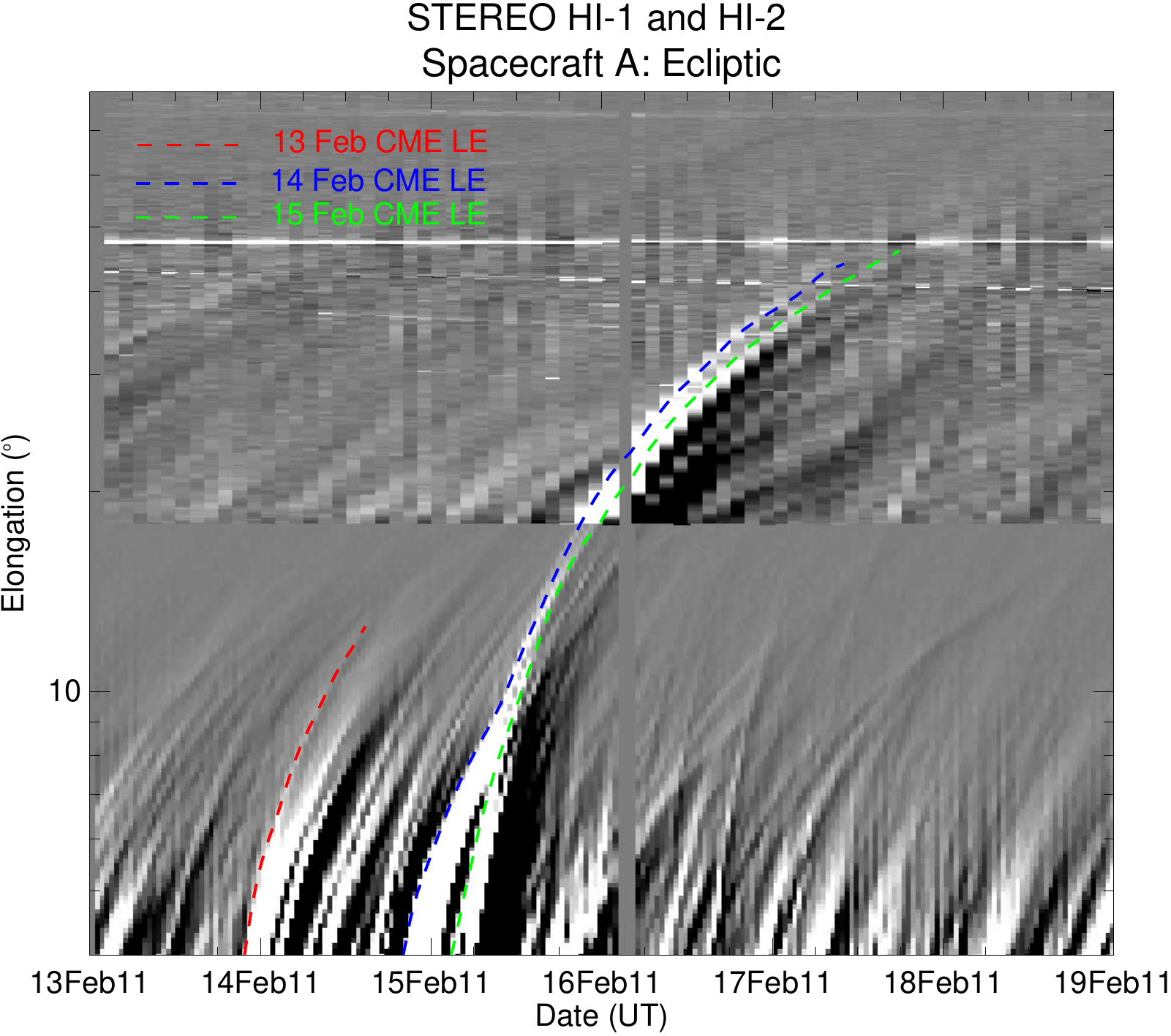}
	\includegraphics[scale=0.5]{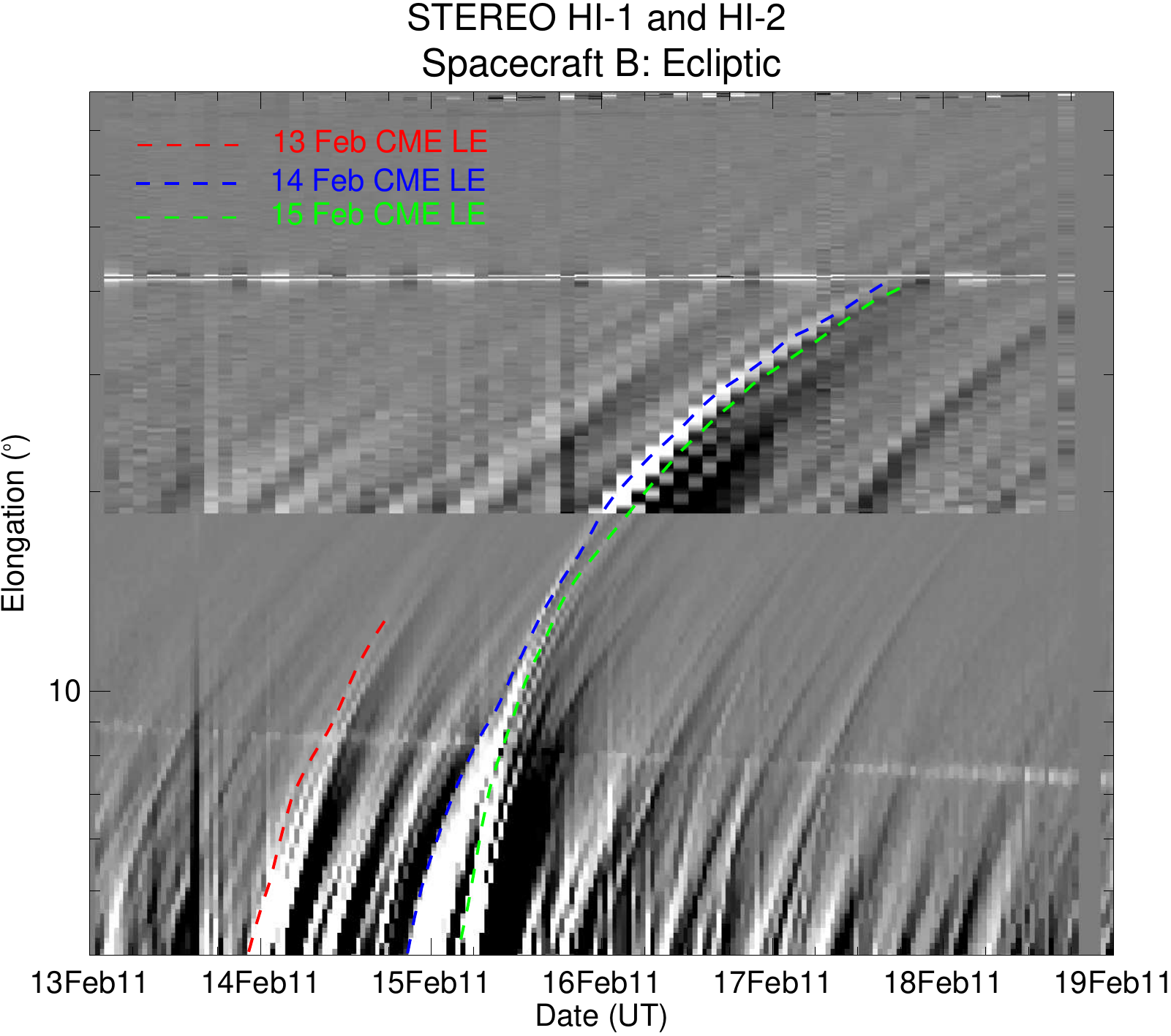}
 \caption{Time-elongation maps (J-maps) for \textit{STEREO-A} (left) and \textit{B} (right) using running differences of images HI1 and HI2 are shown for the interval of 2011 February 13 to 19. The tracks of CME1, CME2 and CME3 are shown with red, blue, and green, respectively.}\label{J-maps}
 \end{figure}

\begin{figure}[h]
 \centering
  \includegraphics[scale=0.80]{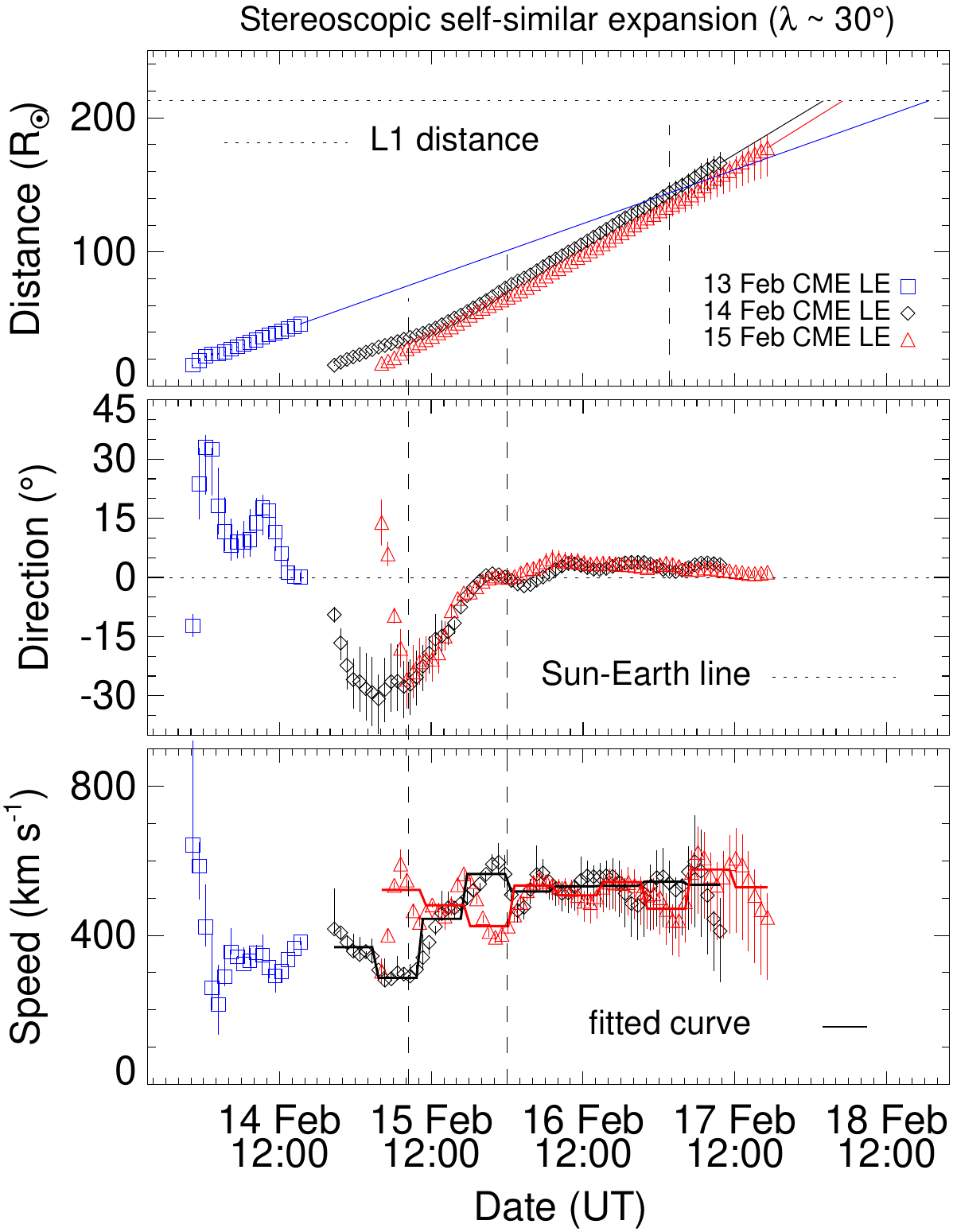}
 \caption{From top to bottom, distance, propagation direction and speed (as obtained using SSSE method) of CME1 (blue), CME2 (black) and CME3 (red) are shown. In the top panel, horizontal dashed line marks the heliocentric distance of L1 point. In the middle panel, dashed horizontal line marks the Sun-Earth line. In the bottom panel, speed shown with symbols are estimated from differentiation of adjacent distances points using three point Lagrange interpolation. The speed shown with solid line is determined by differentiating the fitted first order polynomial for estimated distance for each five hr interval. From the left, first and second vertical dashed lines mark the start and end of collision phase of CME3 and CME2. In the top panel, rightmost vertical dashed line marks the inferred interaction between CME2 and CME1. The vertical solid lines at each data points show the error bars, explained in Section~\ref{CompKinem}.}\label{CME123_SSSE}
 \end{figure}

\begin{figure}[h]
 \centering
  \includegraphics[scale=0.90]{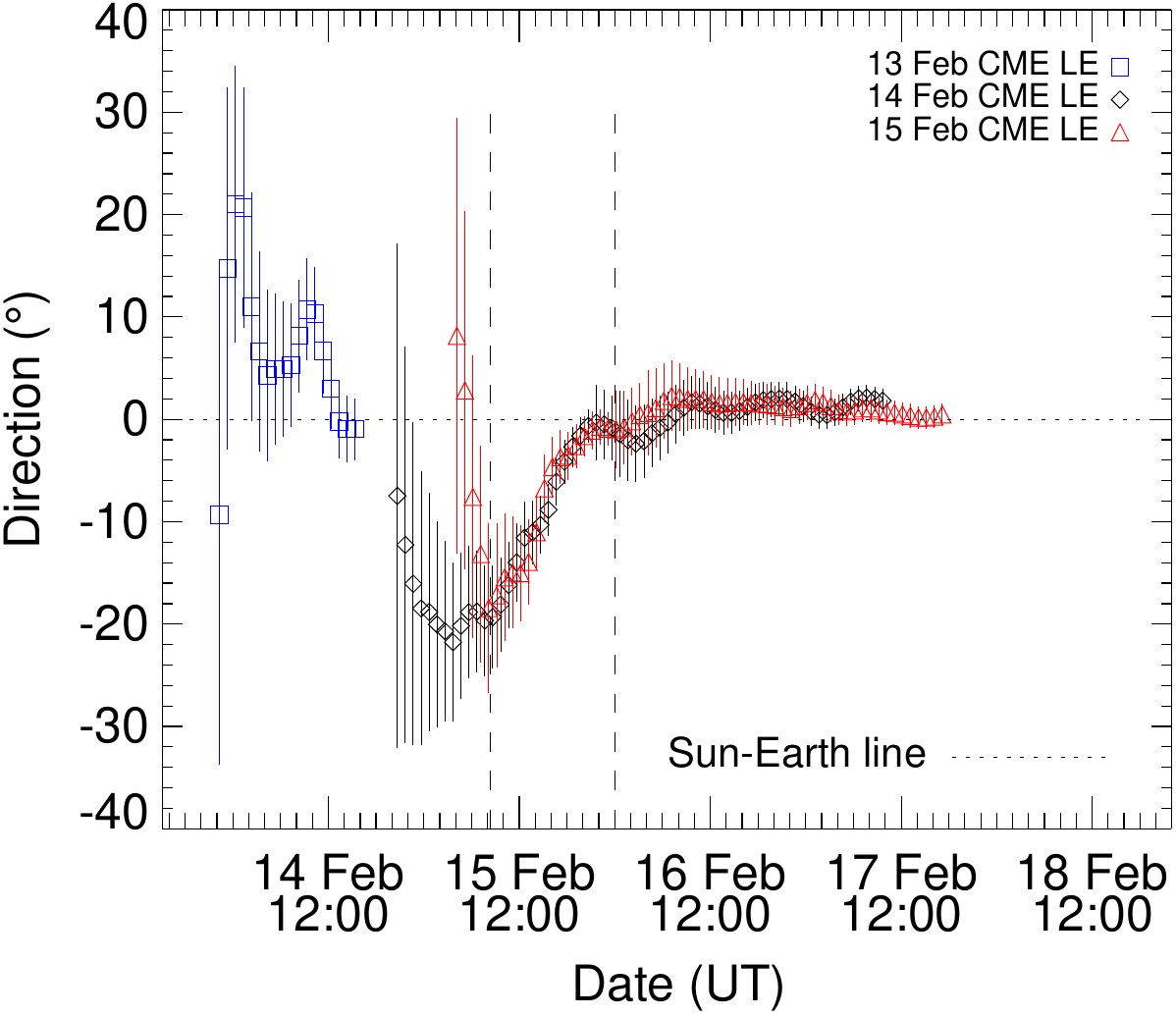}
 \caption{The estimates of direction for CME1 (blue), CME2 (black) and CME3 (red) from GT method is shown. The vertical lines at each data points show the error bars which are calculated mathematically considering the uncertainties of 10 pixels in elongation measurements, i.e., in tracking of CMEs. The two vertical dashed lines mark the start and end of the collision phase of CME2 and 
CME3. The horizontal dashed line denotes the Sun-Earth line. The negative (positive) angle of direction stands for East (West).}\label{CME123_GTErr}
 \end{figure}

\begin{figure}[h]
 \centering
  \includegraphics[scale=0.85]{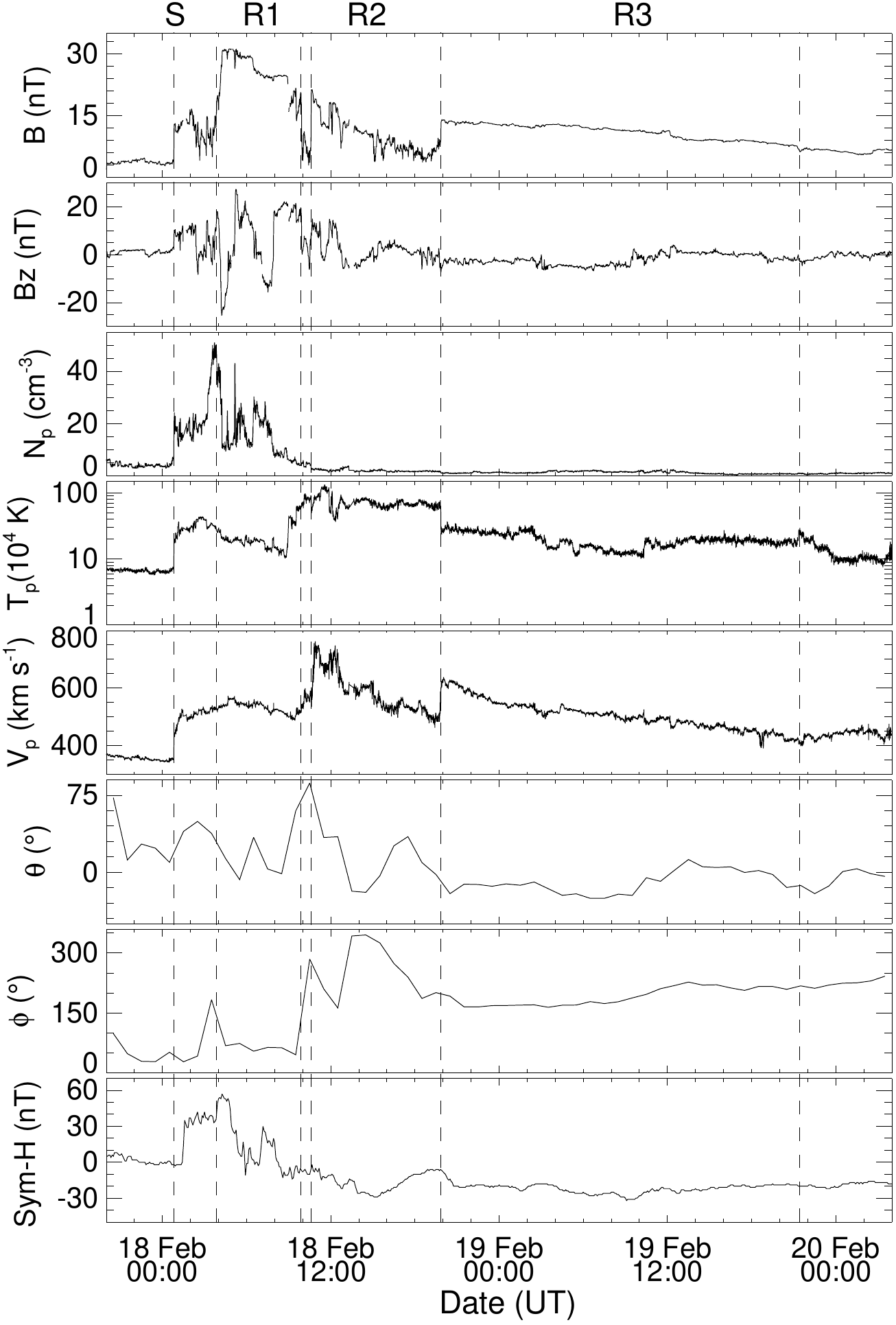}
 \caption{From top to bottom, panels show the variations of magnetic field strength, southward component of magnetic field, proton density, proton temperature, proton flow speed, latitude, longitude of magnetic field vector and longitudinally symmetric disturbance index for horizontal (dipole) direction, respectively. From the left, first, second, third, fourth, fifth and sixth vertical lines mark the arrival of shock, leading edge of CME1, trailing edge of CME1, leading edge of CME2, trailing edge of CME2 and trailing edge of CME3. S, R1, R2 and R3 stand for arrival of shock, bounded interval for CME1, CME2 and CME3 structures, respectively.}\label{CME123_insitu}
 \end{figure}

\begin{figure}[h]
 \centering
  \includegraphics[scale=0.75]{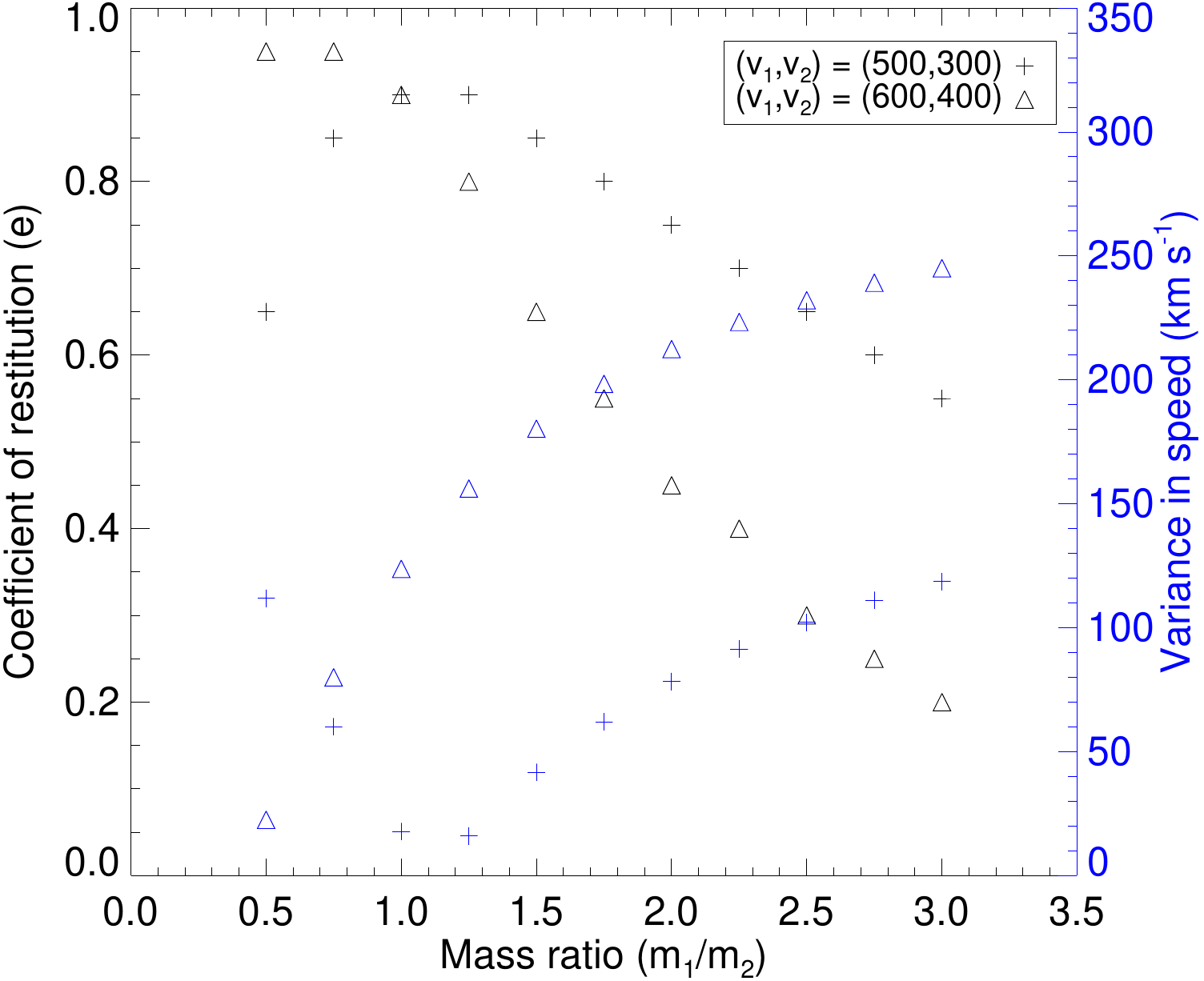}
 \caption{The best suited coefficient of restitution values corresponding to different mass ratios of CME2 and CME3 are shown for  their observed velocity in post-collision phase (in black). We also show the variance in speed corresponding to estimated best suited coefficient of restitution (in blue).}\label{CME12_mass}
 \end{figure}

\end{document}